\documentclass[
 preprint,
 amsmath,amssymb,
 aps,
]{revtex4-1}

\usepackage{subfigure}
\usepackage{graphicx}
\usepackage{dcolumn}
\usepackage{bm}
\usepackage{color}

\begin{document}

\preprint{APS/123-QED}

\title{Super-star networks: Growing optimal\\ scale-free networks via likelihood}

\author{Michael Small}
\email{michael.small@uwa.edu.au}
\author{Yingying Li}
\author{Thomas Stemler}
\author{Kevin Judd}
\affiliation{School of Mathematics and Statistics\\The University of Western Australia, Crawley, WA, Australia, 6009}

\date{\today}

\begin{abstract}
Preferential attachment --- by which new nodes  attach to existing nodes with probability proportional to the existing nodes' degree --- has become the standard growth model for scale-free networks, where the asymptotic probability of a node having degree $k$ is proportional to $k^{-\gamma}$. However, the motivation for this model is entirely {\em ad hoc}. We use exact likelihood arguments and show that the optimal way to build a scale-free network is to attach most new links to nodes of low degree. {Curiously}, this leads to a scale-free networks with a single dominant hub: a star-like structure we call a {\em super-star network}.  Asymptotically, the optimal strategy is to attach each new node to {\em one of the nodes} of degree $k$ with probability proportional to $\frac{1}{N+\zeta(\gamma)(k+1)^\gamma}$ (in a $N$ node network) --- {a stronger bias toward high degree nodes than exhibited by standard preferential attachment. Our algorithm generates optimally} scale-free networks (the super-star networks) as well as randomly sampling the space of all scale-free networks with a given degree exponent $\gamma$. We generate viable realisation with finite $N$ for $1\ll \gamma<2$ as well as $\gamma>2$. We observe an apparently discontinuous transition at $\gamma\approx 2$ between so-called super-star networks and more tree-like realisations. Gradually increasing $\gamma$ further leads to  re-emergence of a super-star hub.  To quantify these structural features we derive a new analytic expression for the expected degree exponent of a pure preferential attachment process, and introduce alternative measures of network entropy. Our approach is generic and may also  be applied to  an arbitrary degree distribution.
\end{abstract}

\pacs{Valid PACS appear here}

\maketitle

\section{Introduction}

Complex networks appear to be virtually ubiquitous \cite{mN10b}, and, moreover, there is currently a growing industry in the study of networks with a power-law distribution of node degree. This recent activity can be traced back to Barab\'asi and Albert's seminal work \cite{aB99} in which they described  the preferential attachment growth model. They showed that if one grows a network by adding nodes in such a way that new nodes preferentially attached to high degree nodes, then the resulting network will be scale-free --- that is, the result of such a process is a connected network with the probability of a vertex having degree $k$ being proportional to $k^{-\gamma}$. 

The surprising power of {\em the Barab\'asi and Albert preferential attachment growth model} (hereafter, simply BA)  is two-fold. First, the scale-free distribution of node degree is implicit and arises naturally. Second, BA generates networks which seem to explain much, but not all, of the great variety of scale-free networks in nature. However, BA does not generate random representative realisations from the set of all scale-free graphs \cite{dC01,aB72}.  In \cite{kJ13} we proposed a Monte-Carlo Markov Chain framework which does just this, thereby showing that many of the properties attributed to a scale-free degree distribution are actually dependent on the particular generative model. 

However, the algorithm in \cite{kJ13} is not a growth model. In this paper we ask: what is the best way to arrive at a scale-free network via a growth process? The surprising result is that BA, is not the right answer.  { The best way of growing a scale-free network is to most often connect to low-degree nodes. This leads to a star-like scale-free network with a single dominant hub: what we call a {\em super-star} network. Careful examination of the relative prevalence of low- and high-degree nodes reveals that, when we cast in a manner analogous to BA, our algorithm chooses nodes with degree $k$ with probability proportional to $\frac{k^\gamma}{N^2}$ (for  $k\ll N$): a {\em stronger bias towards higher degree nodes} than posited in \cite{aB99},} and one that explicitly incorporates the target node degree. { Nonetheless, both our algorithm and BA rely on knowledge of the global degree sequence of the growing network --- yet neither methods needs global connectivity information.}

\begin{figure*}
\[\begin{array}{ccccc}
\includegraphics[width=0.17\textwidth]{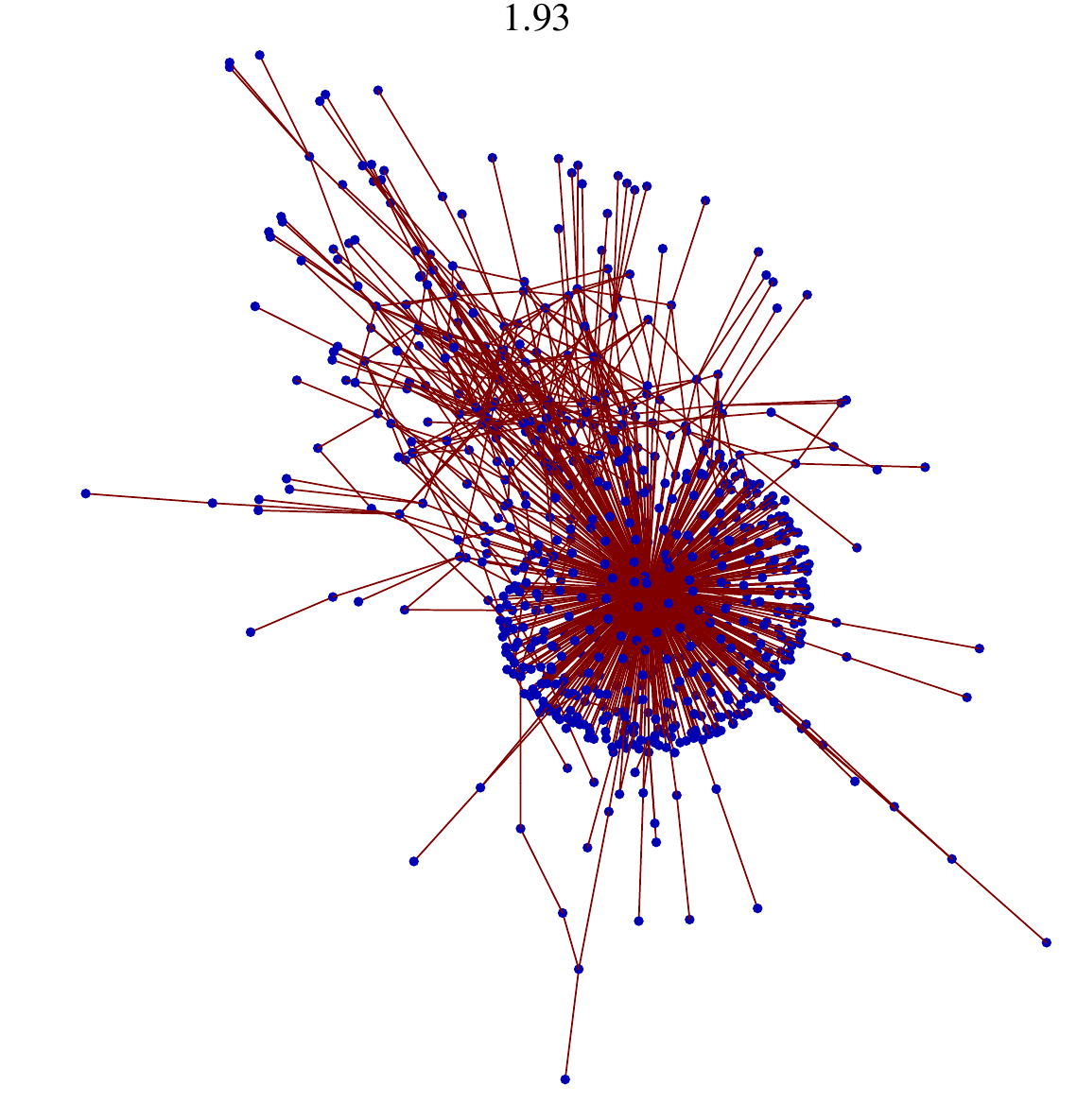} & 
\includegraphics[width=0.17\textwidth]{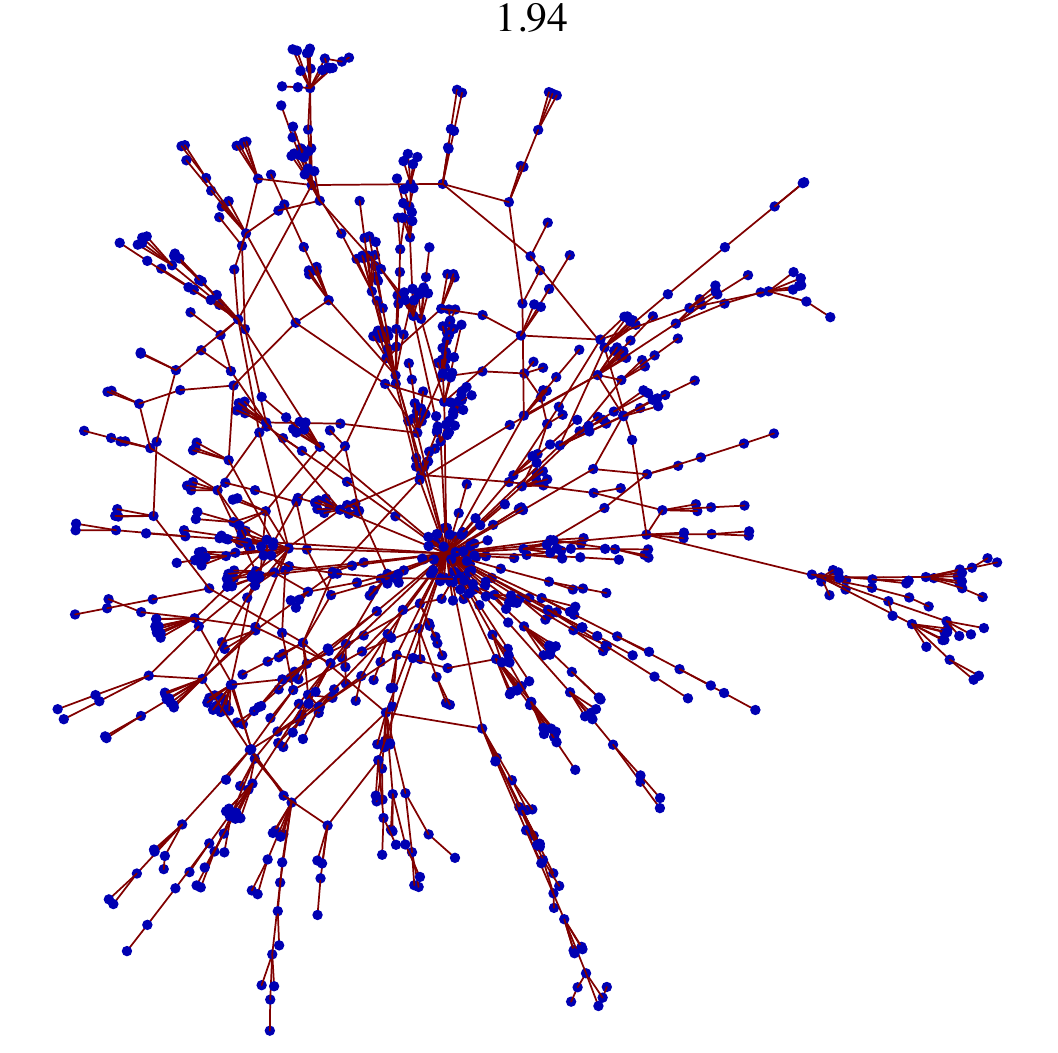} &
\includegraphics[width=0.17\textwidth]{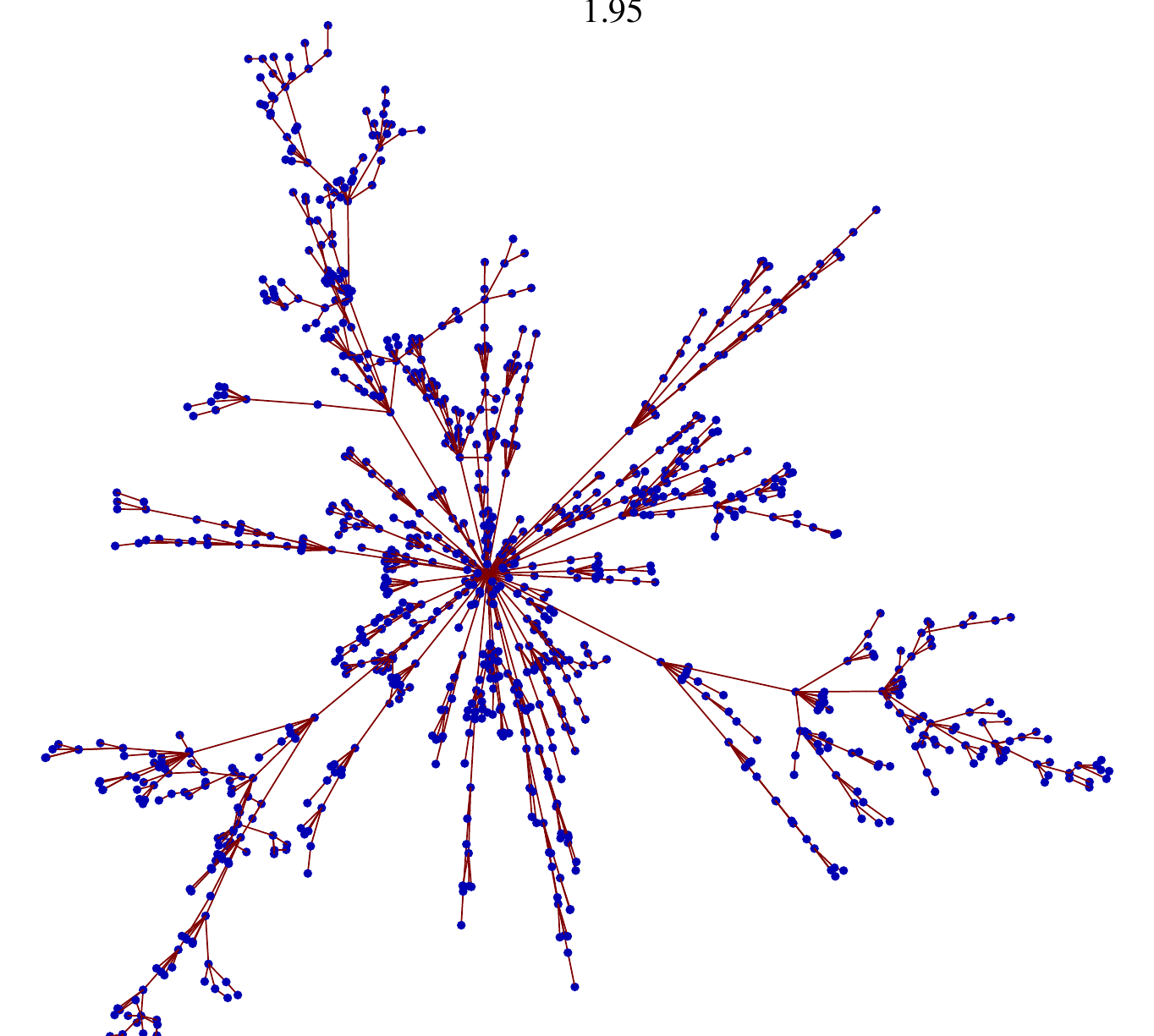} & 
\includegraphics[width=0.17\textwidth]{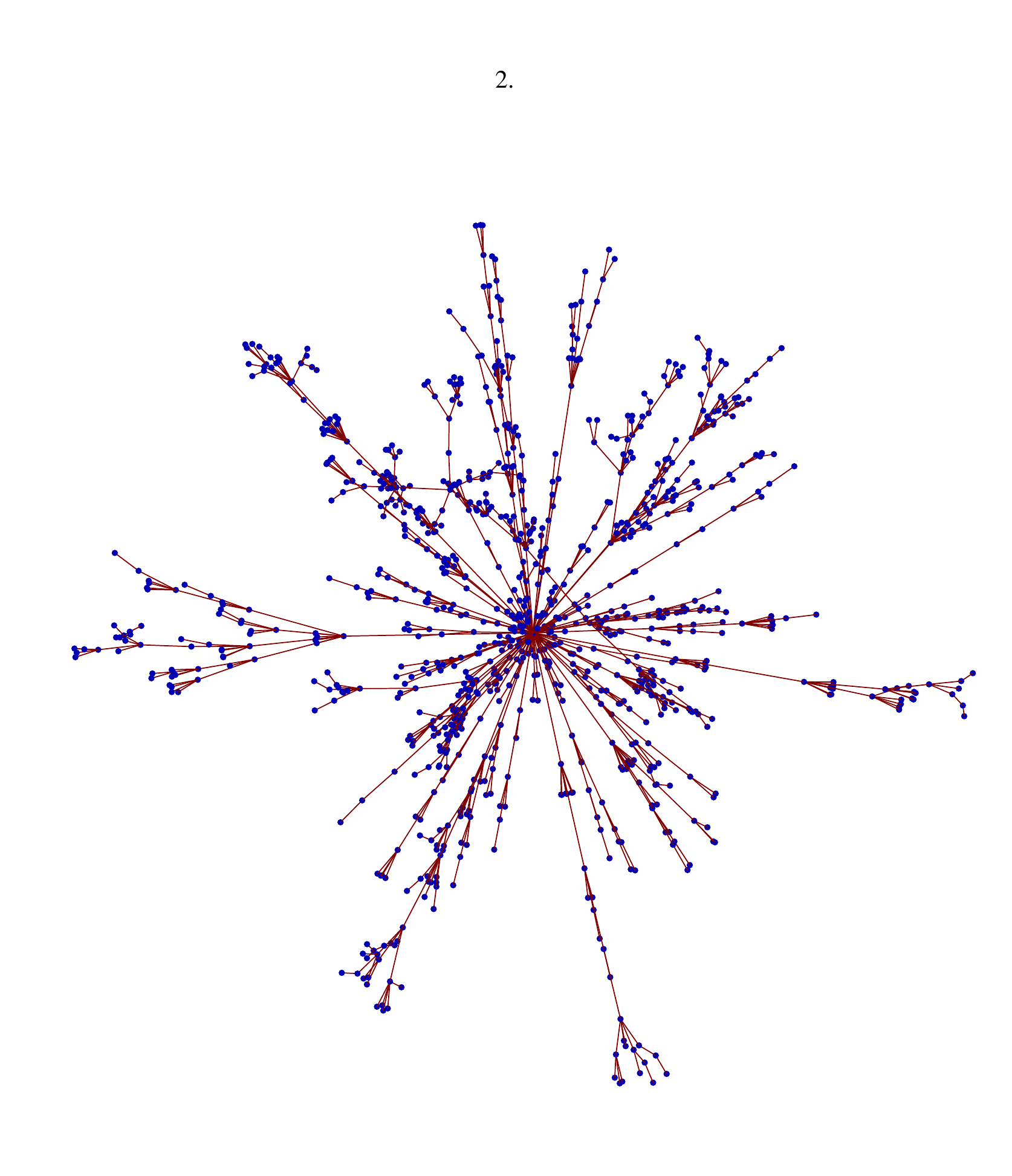} &
\includegraphics[width=0.17\textwidth]{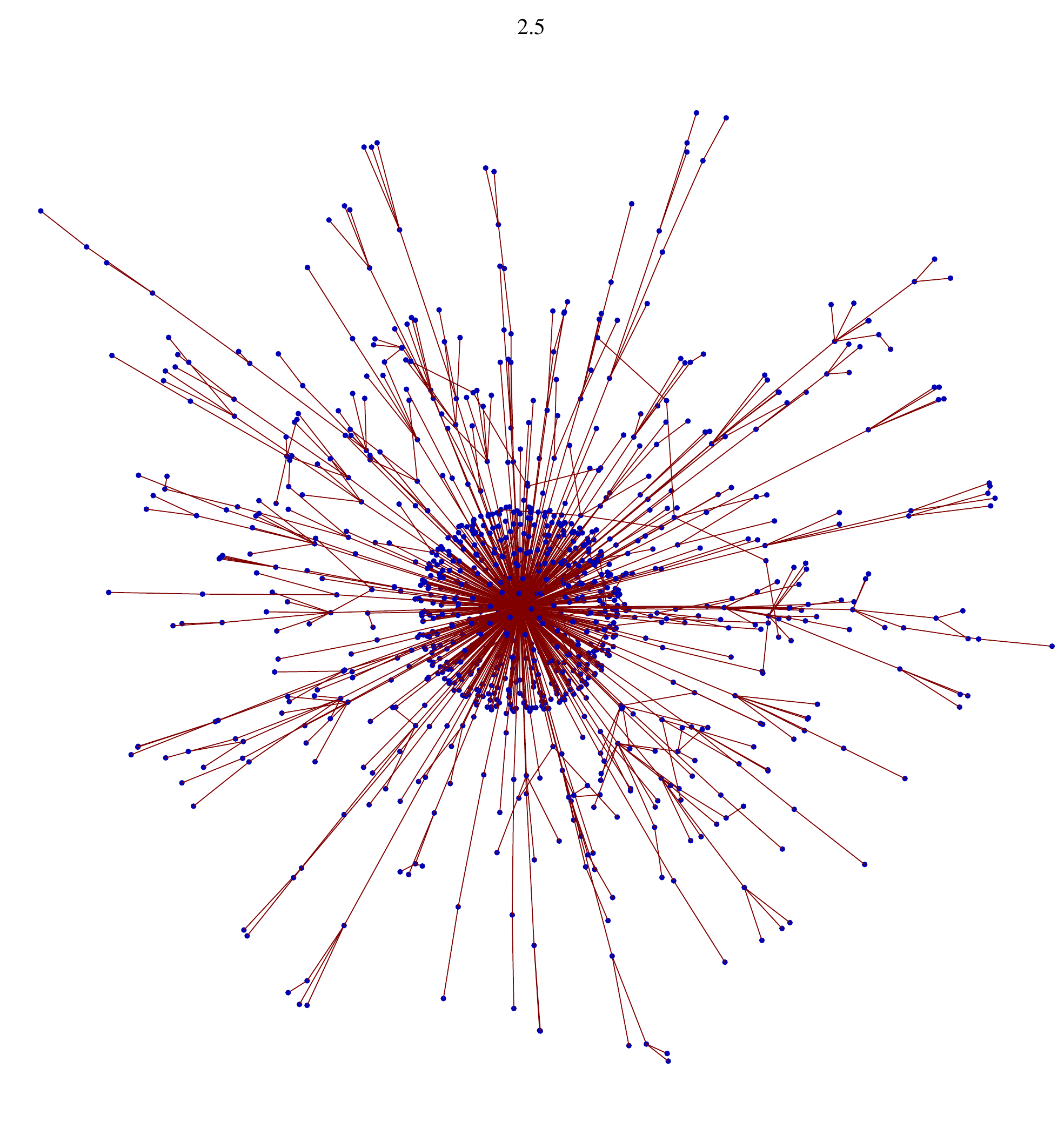}  \\
\gamma=1.93 & \gamma=1.94 & \gamma=1.95 & \gamma=2.0 & \gamma=2.5
\end{array}
\]
\caption{Representative realisations of the optimal scale-free network generation scheme with $\gamma=1.93$, $1.94$, $1.95$, $2.07$ and $2.5$ 
($N=10^3$). For $\gamma\ll 2$ the algorithm is unable to find viable networks, however, for $1\ll\gamma<2$ viable networks do exist (certainly for finite $N$ it is easy to see why this is the case) and such {\em dense} networks are, in-fact, quite plentiful \cite{kJ13}. As $\gamma$ approaches $2.0$ there is a sudden phase transition to large tree-like networks with a proliferation of cross-links.  As $\gamma$ increases further, the cross-links gradually become scarcer and the networks we obtain evolve from highly tree-like back to hub-centric super-star networks.}
\label{egs}
\end{figure*}

Our result does not diminish the remarkable observation that BA will naturally lead to a scale-free network. Conversely,  our algorithm incorporates the scale exponent $\gamma$ explicitly and hence allows one to grow a scale-free network of arbitrary exponent --- indeed the algorithm can easily be modified to grow networks of an arbitrary degree distribution. { Our algorithm provides a mechanism for the generation of scale-free (often non-small world  \cite{birdflu}) networks  with $\gamma<2$.} Such networks have been widely observed \cite{birdflu,rL05,fL01}, but largely viewed as inconsistent with the current generative models { and hence rather pathological}.  Typical networks produced by this algorithm are depicted in Fig. \ref{egs}.

Section \ref{stars} introduces our primary algorithm and an analytic derivation of the optimal (maximum likelihood) model for growth of a scale-free network. In Sec. \ref{cutoff} we address some minor issues concerning the maximum degree of the resultant network --- we extend the results of Sec. \ref{stars} to build truncated power-law networks. Section \ref{conclusion} concludes. In Appendix \ref{egamma} we address some technical issues concerning the dependence of scale-free exponent $\gamma$ on the number of added links $m$ (this extends the well known asymptotic result that in the tail of the distribution $\gamma\rightarrow 3$ independent of $m$ as $N\rightarrow\infty$). Finally, Appendix \ref{gmlg} addresses some technical computational issues concerning growth maximum likelihood networks for $m>1$.
  

\section{Greedy maximum likelihood growth} 
\label{stars}

Let $G_N$ be a network of $N$ nodes. Let $n_k$ denote the number of nodes in $G_N$ with $k$ links. Hence $N=\sum_kn_k$ and ${\bf n}(G_N)=[n_1,n_2,\ldots,n_k,\ldots,n_N]$ is the histogram of degree distribution. { A} scale-free network is  { usually}  {\em defined} to be a graph for which the probability of a node having degree $k$ is given by 
\begin{eqnarray}
\label{powerlaw}
p_k &=& \frac{k^{-\gamma}}{\zeta(\gamma)}.
\end{eqnarray}
The denominator $\zeta(\gamma)$ is the Riemann Zeta function and provides the necessary normalisation. {However, this is a very restricted definition of scale-free --- and it will not be sufficient for our purposes. This equality can only hold asymptotically and for real networks deviation from this definition is to be expected.}  For BA one must also account for the minimum degree $m$, such that  $p_k=0
$ for $k\leq m$ and hence (\ref{powerlaw}) is replaced by a shifted power law. {In what follows we will provide a rigorous probabilistic definition of scale-free --- based on the likelihood of an observed degree definition conforming to the ideal distribution $p_k$.}

Starting from some seed network $G_s$ of $s$ nodes we wish to add nodes so that
\begin{eqnarray}
\label{asymptote}
\lim_{N\rightarrow\infty} \frac{n_k}{N}=  p_k.
\end{eqnarray}
We do this with a series of moves --- adding both nodes and edges --- in such a way that we produce networks that are increasingly likely to satisfy Eq (\ref{asymptote}). The likelihood of $G_N$ conforming to degree distribution $p_k$ is given by the multinomial distribution
\begin{eqnarray}
\label{multinomial}
P(G_N) &=&N!\prod_{k=1}^N\frac{{p_k}^{n_k}}{n_k!}.
\end{eqnarray}
There is a small discrepancy between $n_k$ and $p_k$. As defined above $p_k>0$ for all $k$ --- including $k\geq N$. One way to circumvent this would be to use a truncated power-law instead of (\ref{powerlaw}). However, as this incurs an additional parameter we choose to overlook this at this time. Later we will see that our attachment rule will naturally account for this discrepancy.

Our criteria (\ref{multinomial}) depends only of the node degree --- we assume that all connected networks with a given degree sequence are equally probably \cite{kJ13,linjun1}. One could also condition (\ref{multinomial}) on other desirable network properties: assortativity or embeddability for example. We seek a sequence of moves yielding a sequence of networks with increasing $P(G)$. Each move can either add a new node with one edge, or add another edge from the last added node to the rest of the network. { W}e only propose moves which are modifications to the last node added, and its connections --- we add nodes one at a time and with a variable number of links. 

The first move we consider is the addition of one node of degree one. By adding a new node with one link we move from $G_N$ to $G_{N+1}$ and the marginal payoff  is:
\begin{eqnarray}
Q_{\rm node}(k)  = \frac{P(G_{N+1})}{(N+1)\tilde{p} P(G_N)}
= \frac{(N+1)!\prod_{k=1}^{N}\frac{{p_k}^{n_k'}}{n_k'!}}
{(N+1)\tilde{p}N!\prod_{k=1}^{N}\frac{{p_k}^{n_k}}{n_k!}}
\label{q-node}
&=&\left\{
\begin{array}{cc}
\frac{p_2}{n_2+1}\frac{1}{\tilde{p}} & k=1\\
\frac{p_1}{n_1+1}\frac{n_k}{p_k}\frac{p_{k+1}}{n_{k+1}+1}\frac{1}{\tilde{p}} & k>1
\end{array}\right.
\end{eqnarray}
The additional term $(N+1)\tilde{p}$ accounts for the additional node. To provide a fair comparison between the numerator and denominator we explicitly include the extra node: combinatorially, it could be any of the $(N+1)$ nodes, yet as it is not connected to existing nodes its corresponding probability $\tilde{p}$ is the complement of all other possibilities: $\tilde{p}=1-\sum_{k=1}^Np_k$. { Both simulation and dimensional arguments indicate that this is the correct choice. We find that choice of $\tilde{p}$ which do not scale with $N$ as indicated do not produce viable networks.} Denote by $n_k'$ the terms in the new histogram ${\bf n}(G_{N+1})$. The only difference between the histograms ${\bf n}(G_{N+1})$ and ${\bf n}(G_{N})$ is that $G_{N+1}$ has an additional node with one link (the last node added) and that link connects to a node formerly of degree $k$. Hence: $n'_1 =  n_1+1$;
$n'_k  =  n_k-1$; $n'_{k+1}  =  n_{k+1}+1$; and, $n'_i=n_i$ ({ for} $i\notin\{1,k,k+1\}$).

The second move is the addition of another edge from the last added node to the rest of the network. Suppose that the last node added to the network already has $j$ links (initially, $j=1$). A marginally more complicated counting argument yields:
\begin{eqnarray}
 Q_{{\rm edge-}j}(k)   = \frac{P(\tilde{G}_{N,(j+1)})}{P(G_N)}
\label{q-link}
 &=& \left\{
\begin{array}{cc}
\frac{n_{j-1}}{p_{j-1}}\frac{p_{j+1}}{n_{j+1}+1}  & k=j-1\\
\frac{p_{j+1}^2}{(n_{j+1}+1)(n_{j+1}+2)}\frac{n_j(n_j-1)}{p_j^2} & k=j\\
\frac{n_{j}}{p_{j}}\frac{p_{j+2}}{n_{j+2}+1} &k=j+1\\
\frac{n_j}{p_j}\frac{p_{j+1}}{n_{j+1}+1}\frac{n_k}{p_k}\frac{p_{k+1}}{n_{k+1}+1} & |k-j|>1
\end{array}\right.
\end{eqnarray}
The two cases $|k-j|=1$ are needed to avoid spurious terms when an edge is added between nodes of similar degree.

Equations (\ref{q-node}-\ref{q-link}) are sufficient to define two network growth algorithms --- one a random process, and one a greedy optimisation. We start with a seed network $G_s$ and then at each step we compute $Q_{\rm node}(k)$ and $Q_{{\rm edge}-j}(k)$ as given above. From among these $2N$ marginal payoffs we select the maximum $Q_{\ast}(k)>1$ and adopt the specified move --- either adding a link or a new node.  This  yields a greedy process which  we will call the {\em optimal} scale-free network algorithm (for a given $\gamma$ and $G_s$).  Alternatively, we may seek random realisations by treating each $Q_{\ast}(k)>0$ as being proportional to the probability of accepting that particular move and by selecting among the possibilities randomly. This yields a {\em random} scale-free network growth algorithm(for given $\gamma$). 

Like the BA algorithm, the optimal approach requires knowledge of the degree sequence of the network. Since neither this algorithm nor BA requires true {\em global} knowledge (of clustering, for example), it is easy to imagine generalisation requiring only local information --- utilising the information of a random walker on the network, for example. Of course, this does not address the question of {\em optimality}. We will consider this situation more closely in the future. 

\begin{figure}
\[
\begin{array}{cc}
\includegraphics[width=0.61\textwidth]{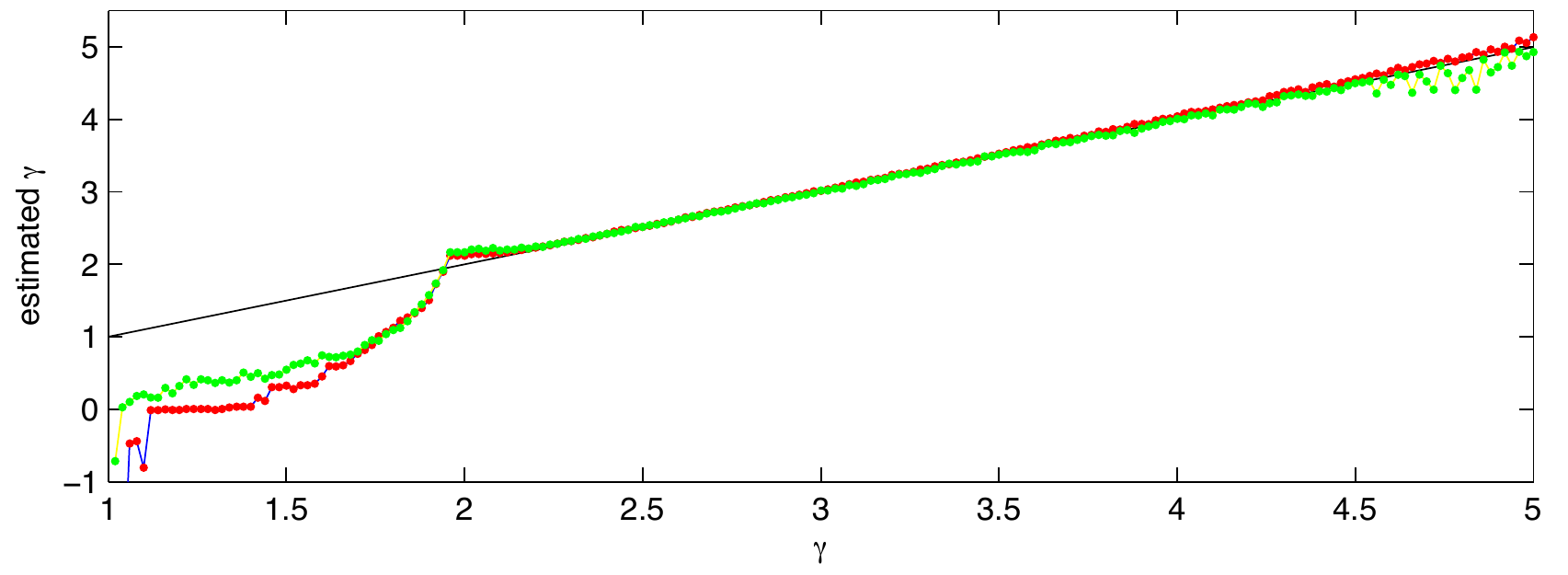}
&\includegraphics[width=0.29\textwidth]{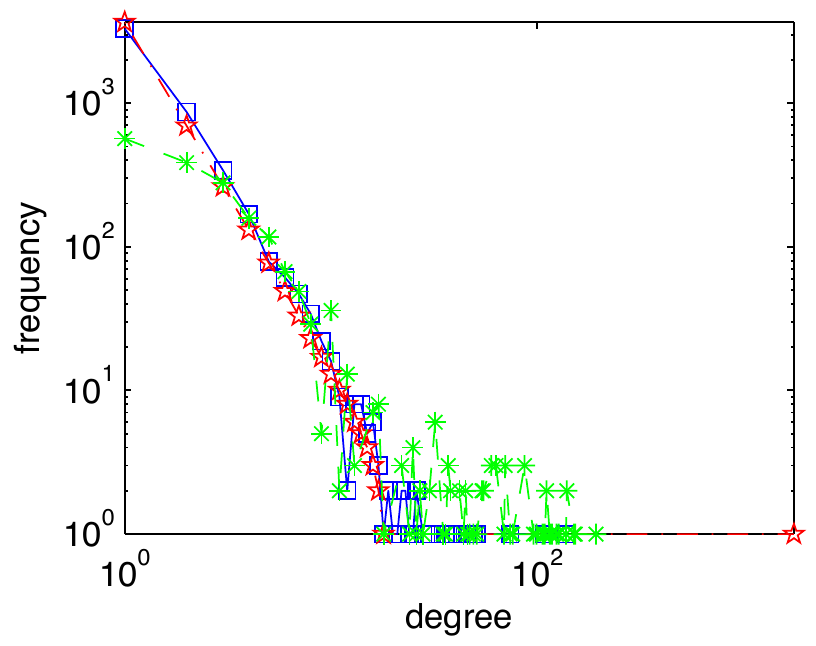}\\
{\rm (a)} &{\rm (b)}
\end{array}
\]
\caption{(a) As we vary $\gamma\in[1.5,5]$ we generate optimal and random realisation of our scale-free network growth algorithm.  We plot $\gamma$ against $\hat\gamma$  (the maximum likelihood estimate of the scale-free exponent from the sample degree distribution) for both optimal (red) and a partially random (green --- slightly larger variance at extrema) realisations with $N=10^4$ nodes. { The identity line is also shown.  (b) typical degree distribution for BA (blue, squares), optimal attachment (red, stars) and semi-random ($q=0.5$, green asterisks).} The partially random scheme { (with $q=0.5$ as described in the text) did} not significantly affect the final degree histogram, and yet this allowed much more variation in the resultant networks { --- f}or $q<1$ we see a transition from super-star networks towards the usual results of BA.}
\label{realgammas}
\end{figure}

In Fig. \ref{egs} we show representative realisations of the optimal scale-free network for $\gamma$ between $1.93$ and $2.5$.  Fig{ure} \ref{egs} depicts the strong dependence of network structure on $\gamma$ and a{n}  explosive transition first to tree-like graphs (at $\gamma\approx 1.94$) and then a gradual drift back toward super-star networks.  Note that $\gamma$ is the degree distribution exponent prescribed for the algorithm, it is not necessarily the actual exponent of the degree distribution of a particular realisation of that algorithm. For comparison, in Fig. \ref{realgammas} we plot exactly this quantity. What we observe is an excellent agreement between the target (prescribed) $\gamma$ and the actual realisations $\hat\gamma$, for $\gamma>2$. For $\gamma<2$ we observe a non-linear relationship (as the degree histogram becomes difficult to satisfy) --- with $\hat\gamma$ systematically smaller than the prescribed value. { Of course, once $\gamma<2$ the mean of the degree distribution is no longer finite (for $N\rightarrow\infty$), nonetheless, such networks do exist (for $N<\infty$) and our algorithm continues to grow them. As the asymptotic results no longer converge, the comparison between $\gamma$ and $\hat\gamma$ is much weaker.}

We now consider the middle ground between our optimal and random algorithms. Define a parameter $q$, such that at each time step there is a probability $q$ of performing an attachment move deterministically, and probability $1-q$ of making a random assignment. The deviation between the values generated by the optimal algorithm { ($q=1$)} and by a somewhat random counter-part ({ $q=0.5$}) is small. We stress that both algorithms only control the degree of the connected nodes --- their {\em location} is always random (one is free to choose from among all nodes of equal degree). The optimal algorithm has a deterministic impact on the histogram, while for the random algorithm both the link placement and  the changes in the histogram are random.  

Next we ask how best to frame these algorithms in terms comparable to BA: we seek to distill, from (\ref{q-node}) and (\ref{q-link}) an analogous rule. Asymptotically, we may suppose (assuming that the algorithm works) that $n_k\rightarrow Np_k $ and hence $n_k\approx \frac{N}{\zeta(\gamma)}k^{-\gamma}$. Substituting this and (\ref{powerlaw}) into (\ref{q-node}) and (\ref{q-link}) gives
\begin{eqnarray}
\label{asymp1}
Q_{\rm node}&\rightarrow&\left\{
\begin{array}{cc}
\frac{N+1}{N +\zeta(\gamma) 2^\gamma} & k=1\\
\frac{N+1}{N+\zeta(\gamma)} \frac{N}{N+\zeta(\gamma)(k+1)^\gamma} & k>1
\end{array}\right.,\\
\nonumber {\rm and}\\
Q_{{\rm edge-}j}  
\label{asymp2}
 &\rightarrow&  \left\{
\begin{array}{cc}
\frac{N}{N+\zeta(\gamma)(j+1)^\gamma}  & k=j-1\\
\frac{N(N-\zeta(\gamma) j^\gamma)}{(N+\zeta(\gamma)(j+1)^\gamma)(N+2\zeta(\gamma)(j+1)^\gamma)} & k=j\\
\frac{N}{N+\zeta(\gamma)(j+2)^\gamma}  &k=j+1\\
\frac{N^2}{(N+\zeta(\gamma)(j+1)^\gamma)(N+\zeta(\gamma)(k+1)^\gamma)} & |k-j|>1
\end{array}\right..
\end{eqnarray}
Hence, we see that the maximal likelihood scale-free network is obtained by observing that the probability of attaching a node to a link of degree $k$ is proportional to 
\begin{eqnarray}
\label{optattach0}
{\rm Prob}({\rm a\ } {\rm degree-}k{\rm\ node}) & \propto & \frac{1}{N+\zeta(\gamma)(k+1)^\gamma}.
\end{eqnarray} 
However, before comparing (\ref{optattach0}) to BA we must note that (\ref{optattach0}) is the probability of attaching to {\em any one of} the nodes of degree $k$ --- asymptotically we expect there to be 
\[n_k\rightarrow Np_k\approx\frac{N}{\zeta(\gamma)}k^{-\gamma}\]
such nodes.  BA  \cite{aB99} says that the probability of attaching to a node, if that node has degree $k$, is proportional to $k$. Hence, our maximum likelihood algorithm approach says that the best thing to do is to link to a node of degree $k$ with probability proportional to
\begin{eqnarray}
\label{optattach1}
{\rm Prob}({\rm node-}i|{\rm degree\ node-}i = k) & \propto & \frac{1}{N}\frac{k^\gamma}{{\frac{N}{\zeta(\gamma)}+(k+1)^\gamma}}.
\end{eqnarray} 
For $k\ll N$ this is proportional to $\frac{k^\gamma}{N^2}$ and as $k\rightarrow N$ this probability is approximately $\left(\frac{k}{k+1}\right)^\gamma\rightarrow 1$ ---  reducing the likelihood of very high degree nodes and acting as an implicit degree cutoff.
While most links are to low degree nodes, we find that high degree nodes are more likely to receive links. The likelihood is significantly stronger than BA --- proportional to an increasing power law --- and dependent on the desired exponent $\gamma$.  The combination $\frac{k^{\gamma}}{N^2}$ reveals an interesting connection between the parameters $\gamma$ and $N$ and the probability $p_k$. Other authors (\cite{sD00,sD00c} for example) have proposed models that are explicitly dependent on the network size $N$, what we do is show that the optimal growth model also has this property.

The difference between (\ref{q-node}-\ref{q-link}) and (\ref{asymp1}-\ref{asymp2}) is that the former incorporates deviation from the target degree distribution $p_k$ as a sequence of ratios of the form $\frac{n_k}{p_k}$. The later assigns links proportional to (\ref{optattach1})   or (\ref{asymp1}-\ref{asymp2}). Unfortunately, an attachment algorithm based on these {\em asymptotic attachment} rules does not perform well. Simulation has shown us that for small $N$ the asymptotic approximation is usually poor, and there appears to be little hope of appropriate convergence.  We could be tempted to define a third implementation of our growth algorithm: randomly select nodes according to the distributions (\ref{asymp1}-\ref{asymp2}) and iterate. Such an algorithm generates what we refer to as {\em asymptotically} scale-free networks.  However, systematic results from this algorithm are not reported here --- and performance is generally very poor. For small seed networks, the initial configuration is far from the asymptotic ideal and convergence is poor. For large seed networks, the result depend on the choice of initial network. 

 \begin{figure}
\[
\includegraphics[width=0.9\textwidth]{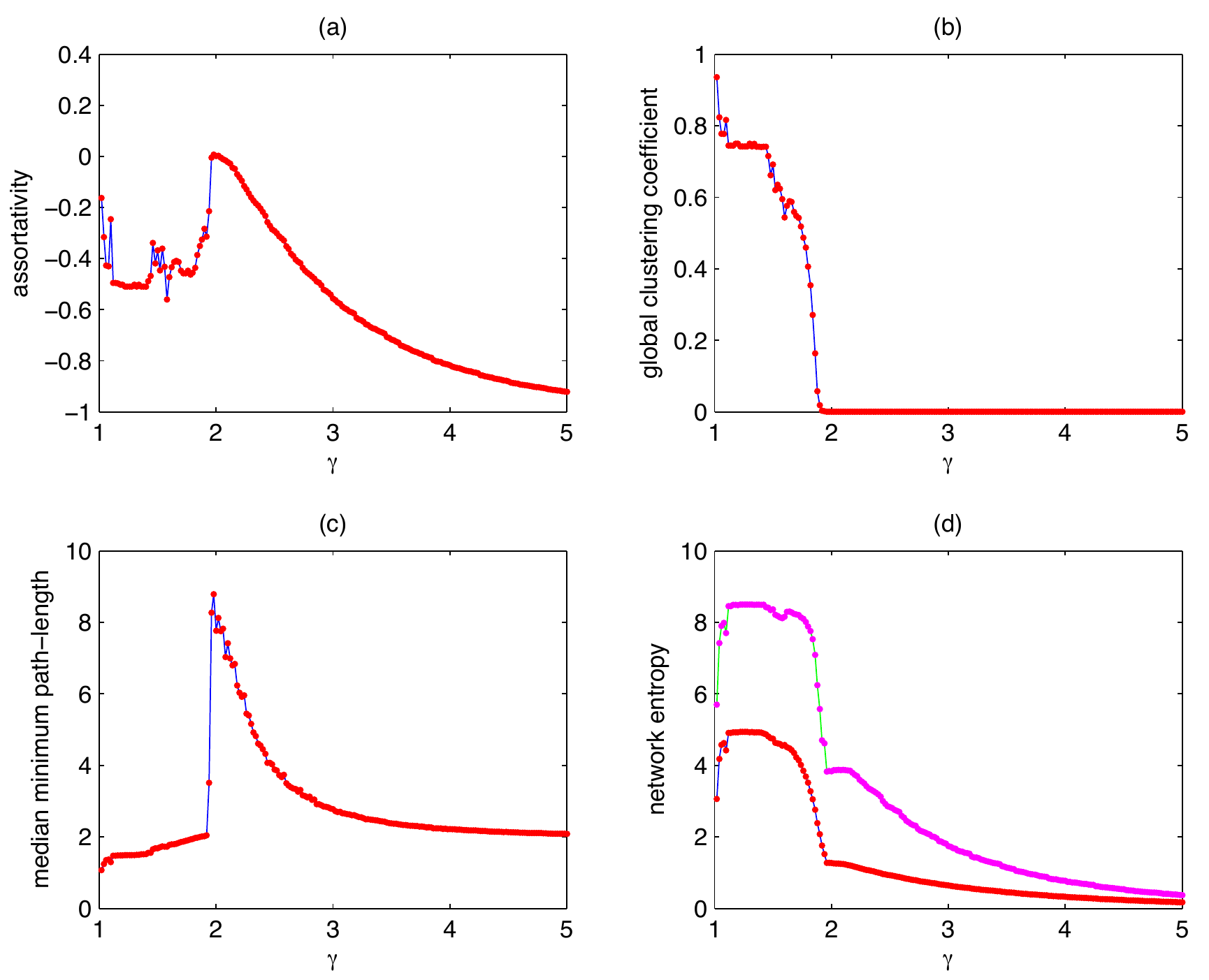} 
\]
\caption{For $\gamma\in (1,5]$ we compute optimal realisations of our scale-free network generation algorithms ($N=10^4$). We show (a) assortativity (linear Pearson correlation coefficient), (b) global clustering coefficient,  (c)  shortest path-length, and (d) two measures of network entropy. Note that the hub-like nature of networks with large $\gamma$ is evident from the  shortest path-length, while global clustering coefficient  drops to zero at $\gamma\approx 2$. Interestingly, assortativity remains negative, peaking with a value of $0$ at $\gamma\approx 2$ and then declines rapidly. The  shortest path-length has clear evidence of the under-size networks for $\gamma\ll 2$, an abrupt transition to tree-like networks near $2$, and then a gradual decay to a single dominant hub as $\gamma$ increases further. The two network entropy statistics  compute the entropy of the degree sequence \cite{gB08,gB09} (lower, red/blue line) and the entropy of the link-degree coincidences (upper, magenta/green line). Both entropy measures show an abrupt, non-differentiable transition at $\gamma\approx2$. Each data point reported in these figures is the corresponding statistic values estimated from a single network (with $\gamma$ increasing between simulations by $0.01$), variance can be inferred from the smoothness of the plotted curves.}
\label{props}
\end{figure}

In Fig. \ref{props} we compute the usually quoted properties of scale-free networks, for realisations of our first two algorithms. We observe a systematic dependence of these properties on the exponent $\gamma$. In particular, there is a sudden --- apparently not differentiable --- transition in the vicinity of $\gamma\approx 2$ as the network structure rearranges. This transition is indicative of the underlying structural change in the network structure for $\gamma$ bigger than two --- the onset of the super-star hub structure depicted in Fig. \ref{egs}. We note that for $\gamma>2$ the typical maximal for the connectivity scales as
$N^{1/(\gamma-1)}$ \ref{mB04}. For $\gamma<2$ this should grows faster
than linearly resulting in the (low-degrees exponent) emergence of a superstar structure
in our model.

\begin{figure}
\[
\begin{array}{c}
\includegraphics[width=0.9\textwidth]{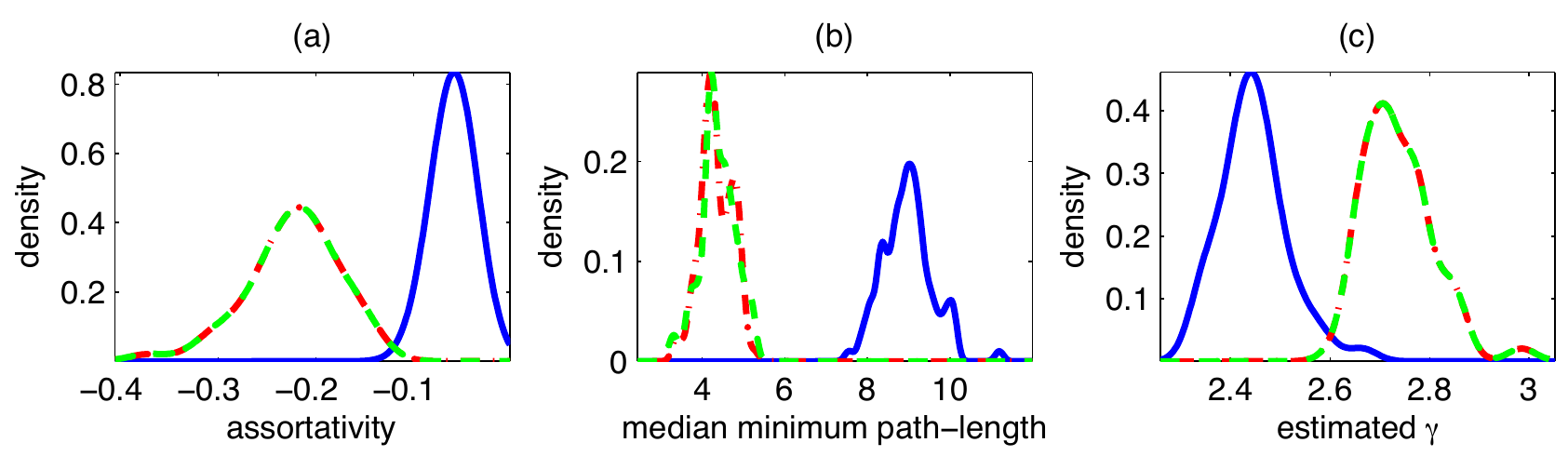} \\
\includegraphics[width=0.9\textwidth]{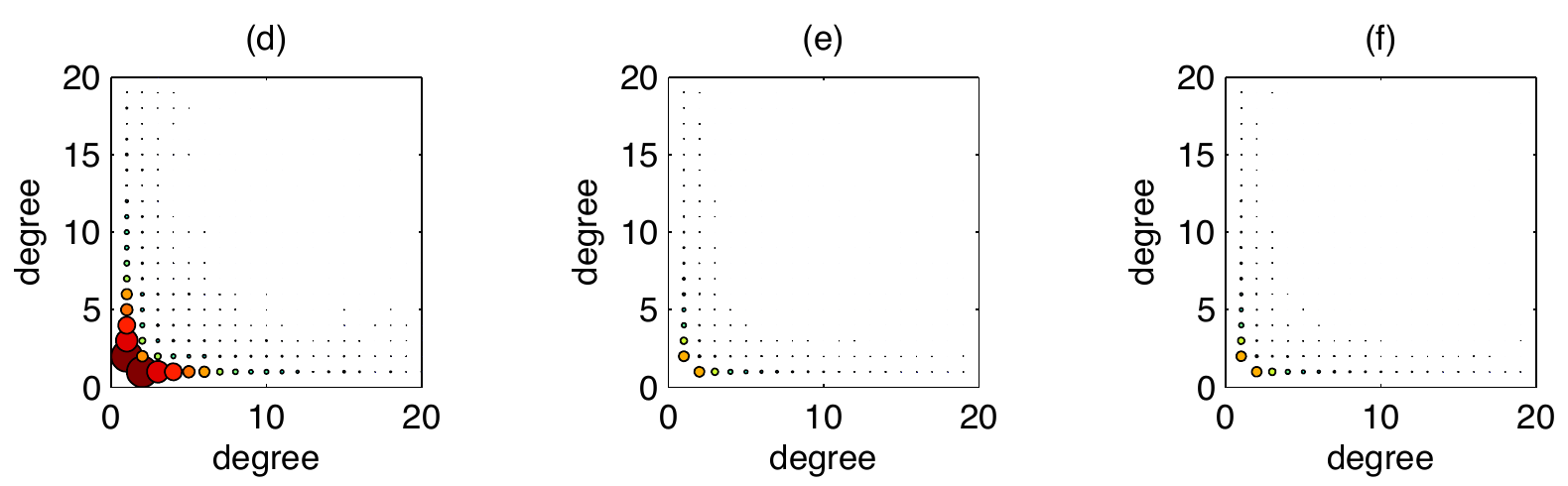} 
\end{array}
\]
\caption{We generate BA networks (blue solid lines), estimate the scale exponent $\gamma$, and then generate networks according to our first two schemes with this estimated value $\hat\gamma$ and $N=10^4$ nodes (red dot-dashed and green dashed lines). Displayed here are the usual network properties estimated from the resultant networks and depicted as histograms (generated from $100$ network realisations via a Gaussian kernel smoothing algorithm): (a) assortativity, (b) shortest path-length, and (c) estimated exponent $\hat\gamma$ { (adaptive binning)}. Remarkably, the distributions reported for the optimal scheme (red dot-dashed) is almost identical to the results of the randomisation scheme ($q=0.5$  and green dashed lines). As assortativity is a linear measure, it is not particularly good at describing the detail of degree-degree correlation. Panels (d-f) illustrate scatter plots (circles size and colour proportional to likelihood/number) of actual degree-degree structure for representative networks: (d) BA; (e) optimal; and, (f) random.  }
\label{newfig}
\end{figure}

More generally, we find the range of behaviours demonstrated in Fig. \ref {props} is far wider than what one would observe with straight-forward BA. In addition to network topological measures, we also report two measures of network entropy. First, following Bianconi \cite{gB08,gB09} (and our own independent and {\em ad hoc} treatment \cite{iscas13}), we compute the entropy of the events defined by the sample degree histogram. That is, $-\sum \frac{n_k}{N} \log \frac{n_k}{N}$. However, this quantity does not take into account the structure between nodes of the network (i.e. how the high and low degree nodes are distributed within that topology). Hence, we also compute what we call the {\em network link entropy}. If $e_{i,j}$ is the sample probability of an edge joining nodes of degree $i$ and $j$ the this version of entropy is computed as $-\sum_{i,j} e_{i,j}\log{e_{i,j}}$.  Fig. \ref{props} (d) illustrates the result of both computations.

Figure \ref{newfig}  demonstrates that, even if $\gamma$ is restricted to values which one obtains from BA, the schemes we propose here exhibit a much wider variation in network structures. {T}he assortativity of our algorithm is stronger (more negative), and, as  consequence the mean path-length is lower. { The BA algorithm illustrated here has minimum degree $m=1$ and hence one can compute (see Appendix \ref{egamma}) that the expected value of $\gamma$ is $\approx 2.471$, in excellent agreement with Fig. \ref{newfig} (c). In comparison, the optimal algorithm adds more links per node and achieves a significantly higher value of $\gamma$.} 


\section{Truncated power-laws, an upper bound on node degree and a ``natural'' cutoff function}
\label{cutoff}

In the previous section we noted that Eqn (\ref{optattach1}) acted as an implicit cutoff and limits the growth of very high degree nodes. While it is pleasing to observe the manifestation of this cutoff directly in the maximum likelihood model, there are several models that impose explicit (but so-called ``natural'') cutoffs in the power-law degree distribution of evolving scale-free networks \cite{mB04}. It is perhaps useful to ask whether the addition of a explicit mechanistic cutoff to our optimal scheme will either significantly alter our results or provide an explanation for the super-star effect and deviation from the BA model. Hence, in this section we explore the effect of imposing an arbitrary maximum degree $C_{\rm max}<N$ on  a scale-free network network following the optimal growth procedure described in Sec. \ref{stars}. 

The optimal method proposed in the previous section tends to generate scale-free networks with a single big hub, which means the degree distribution in log-log scale has one particular extremum in the far tail of the distribution far away from the power-law. Because this datum corresponds to an extremely low-probability event it does not significantly effect the likelihood evaluation and is hence a natural explanation. Conversely, we note that the BA method produces networks with significant deviation amount the low degree nodes --- deviation which is explicitly typically ignored when estimating the exponent of such networks (see App. \ref{edeg}). Nonetheless, our networks produce a small deviation from the ideal scale-free property in the resulting degree histogram and it is natural to wonder whether the super-star networks we observe are only an artefact of this single node in the far tail of the distribution. To test this we repeat the optimal growth process described in Sec. {\ref{stars} after modifying (\ref{powerlaw}) to include an explicit cutoff function. We now consider the truncated power-law distribution $p_k$.
\begin{eqnarray}\label{equ::1}
p_k &= & \left\{ 
\begin{array}{cc}
Ck^{-\gamma}, &{k\leq C_{\rm max}}\\
Ck^{-\alpha \gamma},&\text{otherwise}
\end{array}\right.
\end{eqnarray}
where $C$ is the normalisation constant and $\alpha$ can be any number more than 1 when $\gamma > 2$, and $C_{\rm max}$ is the cutoff value.
We call this the {\em truncated} maximum likelihood method --- as this approach is parameterised by the constant $C_{\rm max}$ we will equivalent refer to this as the $C_{\rm max}$-method.

When $C_{\rm max}$ is large, the effect of the additional cutoff term in the power-law distribution function is small, and the truncated method behaves similarly to the greedy optimal method described above: the network results in a single dominant hub and a potential super-star network. However, as we decrease $C_{\rm max}$, the  cutoff function has more influence and we observe that this gradually splits the super hub to many smaller (but still large) hubs. Figure $\ref{figs/fig5}$ demonstrates this process. That is, the effect of the introduction of a maximum degree $C_{\rm max}$ results in the largest hub being reduced to a rich-club of multiple large (but nonetheless, {\em smaller}) degree hub nodes --- and these nodes are interconnected. One can view this rich club of high degree nodes as a virtual super-hub: replacing these nodes with a single node of degree greater than $C_{\rm max}$ results in networks exactly equivalent to the previous section. The super-star networks and their hub nodes are not an artefact. 

\begin{figure*}[!t]
\centering
\subfigure[\;\;BA] {\includegraphics[width=3.5 cm]{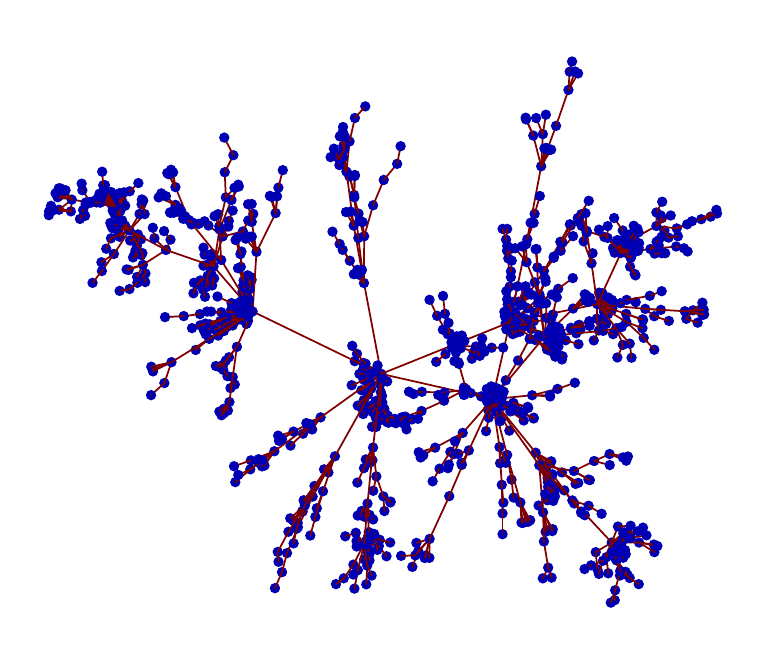}}
\subfigure[\;\;$C_{\rm max} = 20$] {\includegraphics[width=3.5 cm]{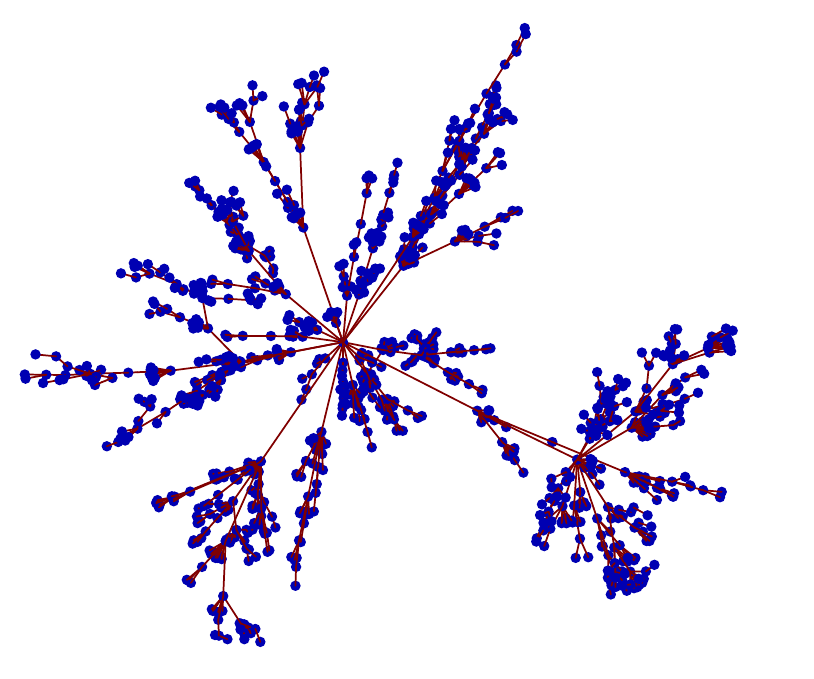}}
\subfigure[\;\;$C_{\rm max} = 100$] {\includegraphics[width=3.5 cm]{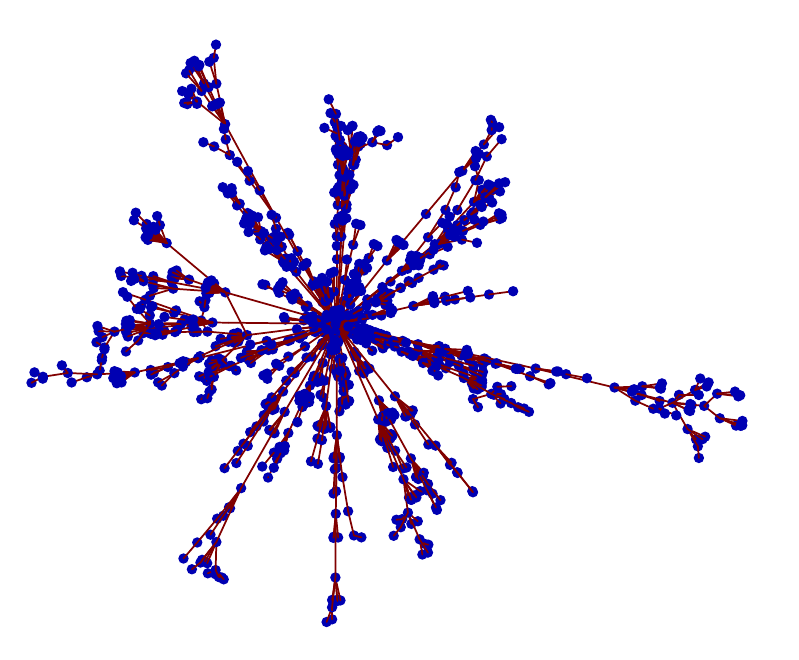}}
\subfigure[\;\;$C_{\rm max} = 200$] {\includegraphics[width=3.5 cm]{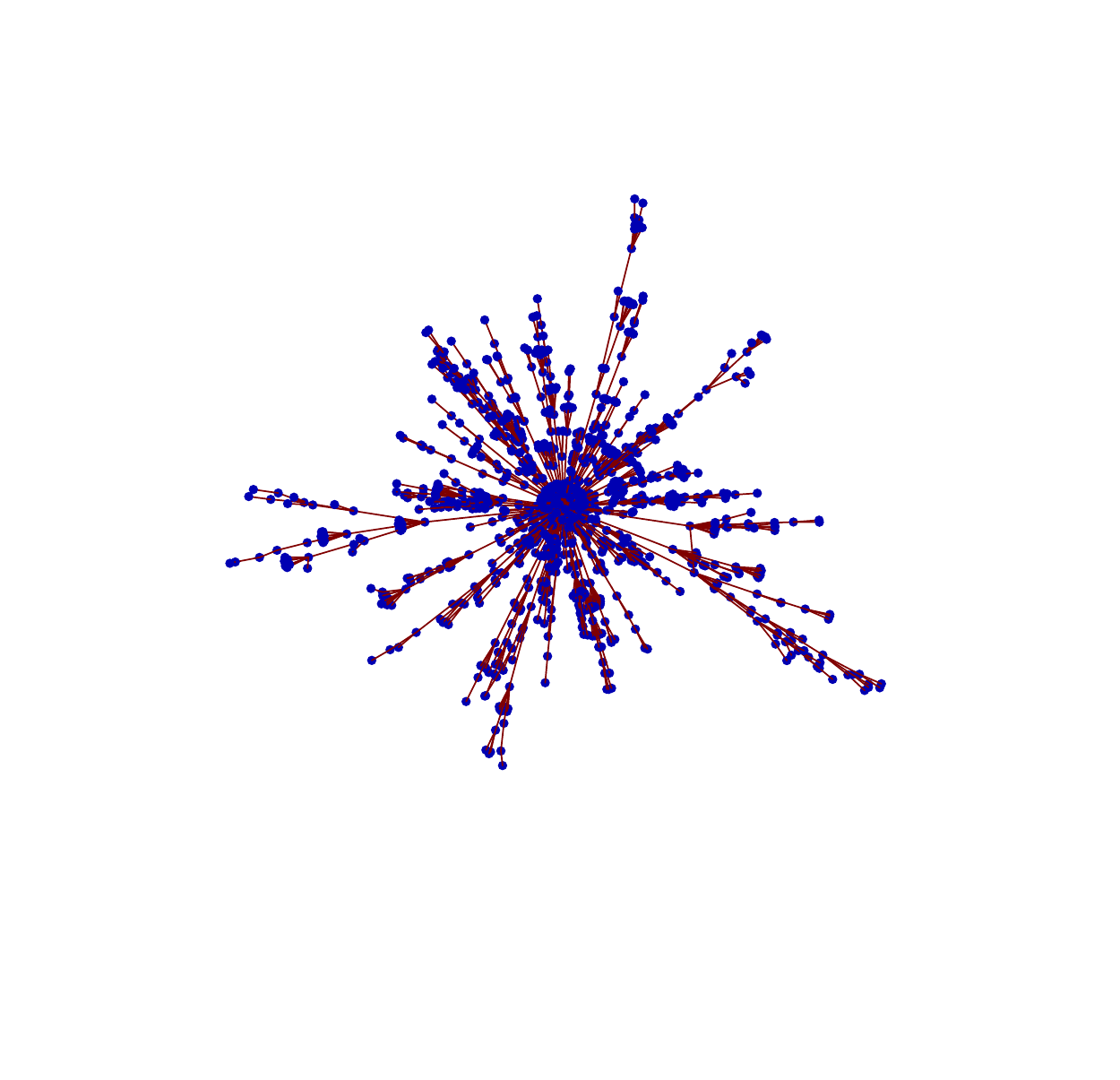}}\\
\subfigure[\;\;$C_{\rm max} = 300$] {\includegraphics[width=3.5 cm]{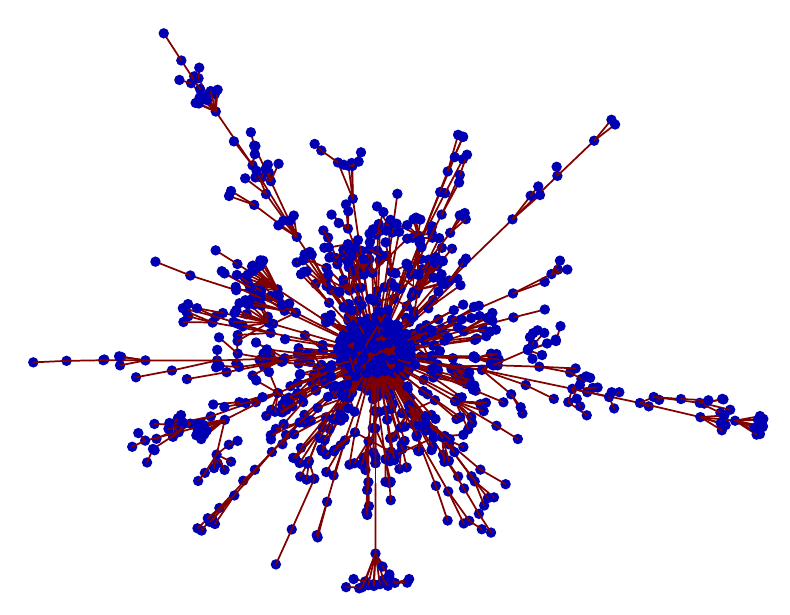}}
\subfigure[\;\;$C_{\rm max} = 400$] {\includegraphics[width=3.5 cm]{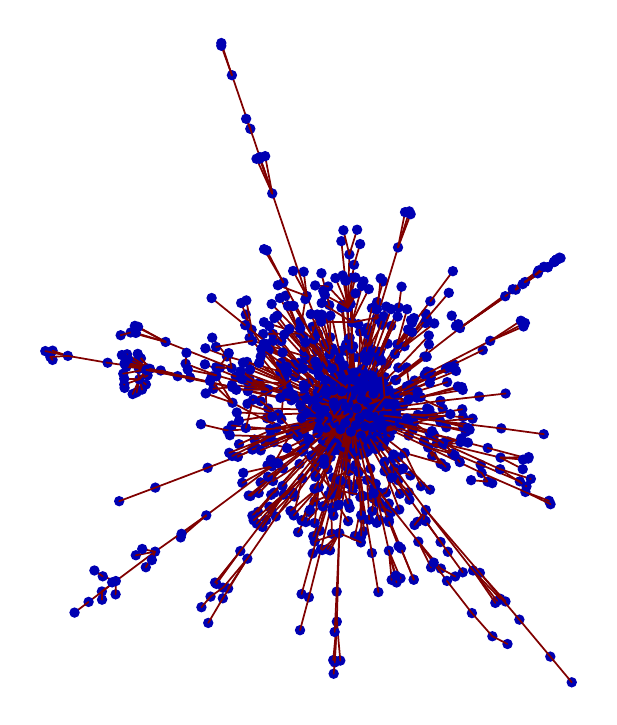}}
\subfigure[\;\;$C_{\rm max} = 900$] {\includegraphics[width=3.5 cm]{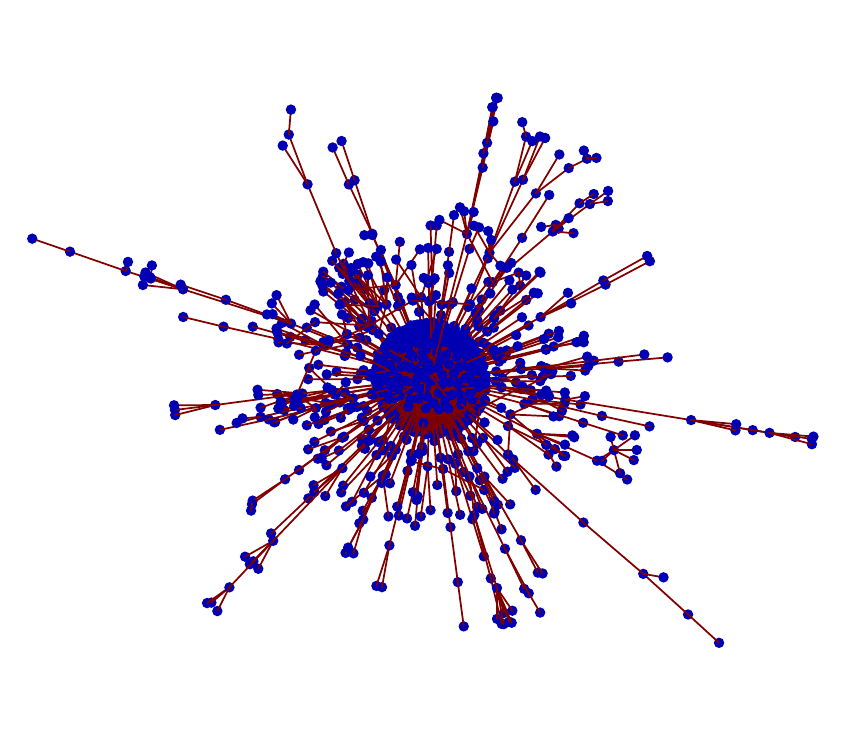}}
\subfigure[\;\;optimal] {\includegraphics[width=3.5 cm]{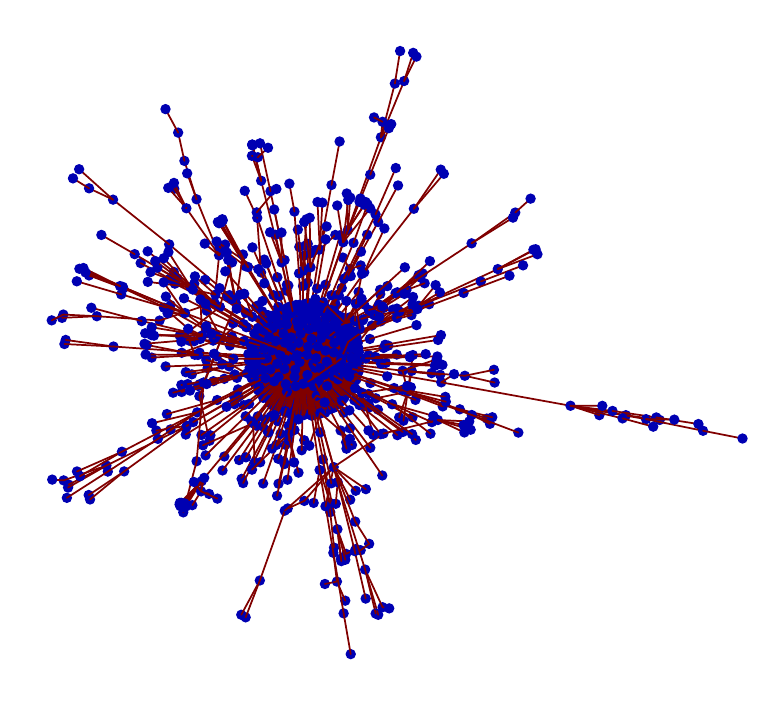}}
\caption{Representative realisations of the BA method, optimal method (Sec. \ref{stars}), and an imposed $C_{\rm max}$  with $C_{\rm max}=20, 100, 200, 300, 400, 900$ ($N=10^3$) (Sec. \ref{cutoff}). When $C_{\rm max}$ is quite small, i.e., $C_{\rm max} \leq 100$, networks look like the BA network because of a scattered structure. For $C_{\rm max} >100$, networks tend to have less hubs and gradually evolve to concentrated super-star networks --- equivalent to the optimal method.}
\label{figs/fig5}
\end{figure*}

\begin{figure*}
\centering
\subfigure[\;\;Assortativity] {\includegraphics[width=5 cm]{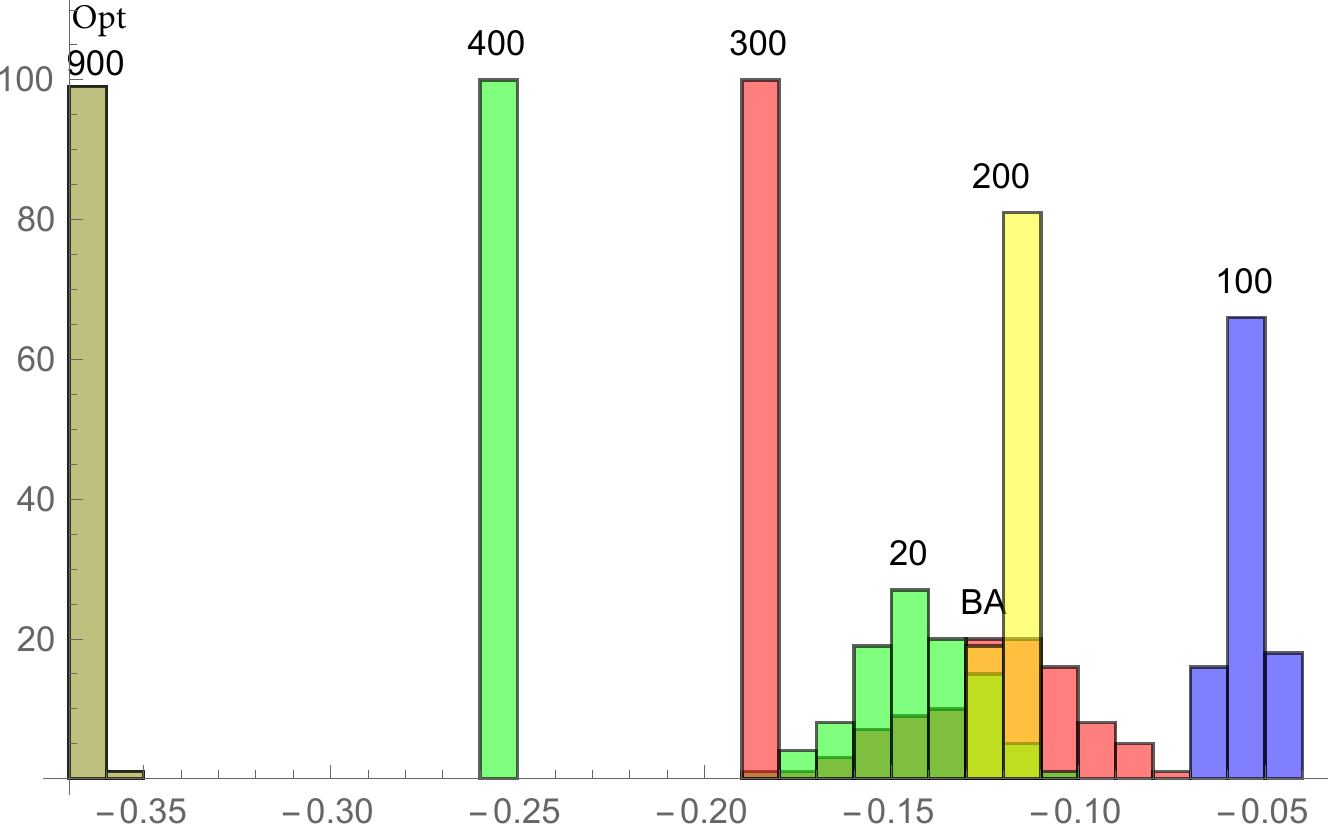}}
\subfigure[\;\;Mean shortest path length] {\includegraphics[width=5 cm]{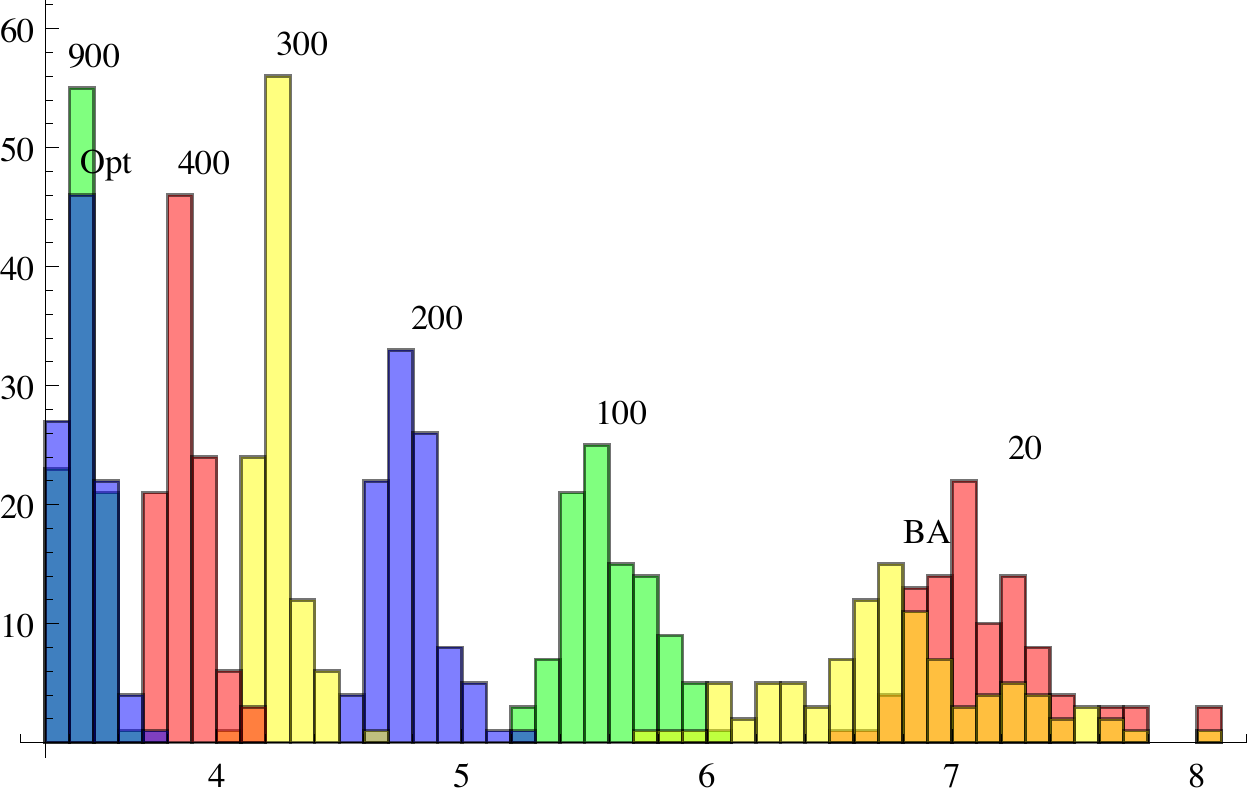}}
\subfigure[\;\;Relative probability] {\includegraphics[width=5 cm]{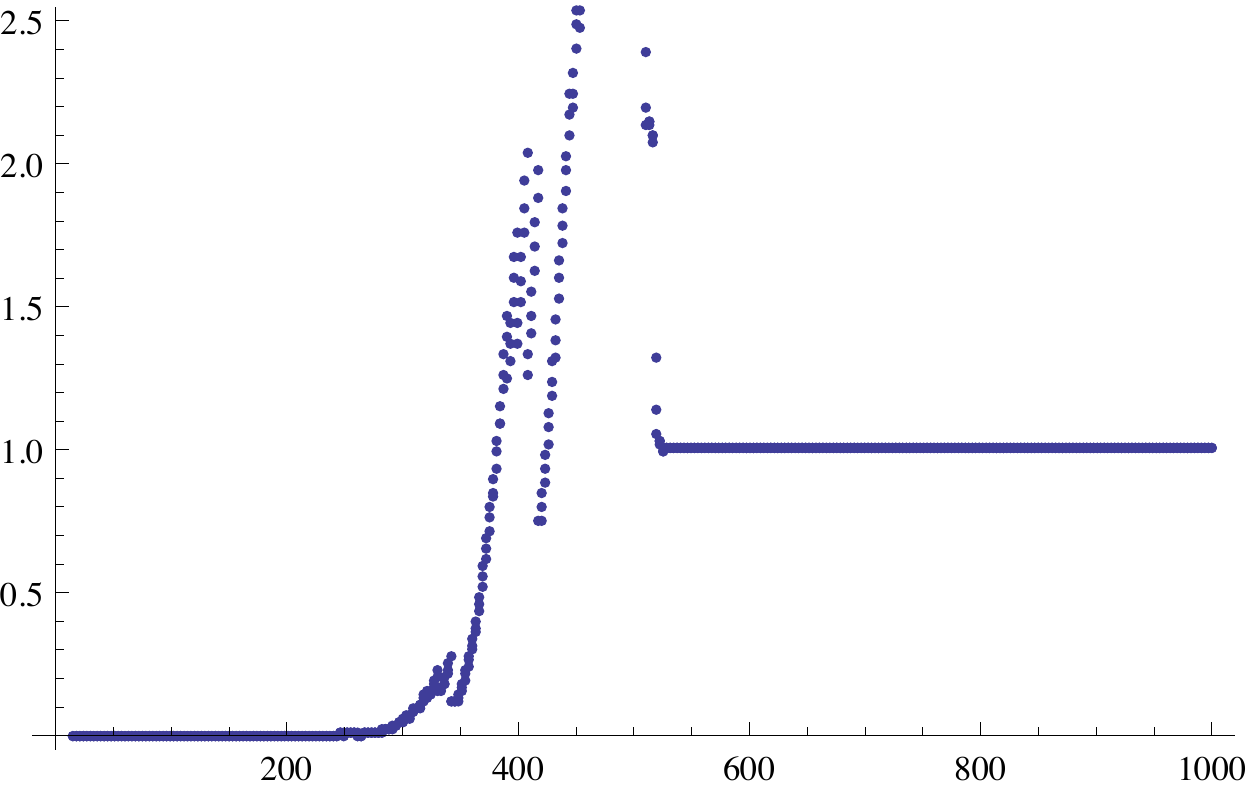}}
\caption{In (a) and (b), we apply the optimal method, the BA method and the truncated $C_{\rm max}$ method with different $C_{\rm max}$ to generate scale-free networks. For each method or each value of $C_{\rm max}$, we generate $100$ networks, and then display the histogram of common properties of scale-free networks: assortativity and shortest path length. The number on the top of the bars indicate the $C_{\rm max}$ value. From the panel (a) and (b) we can see that when $C_{\rm max}$ value is large, the networks we get have very similar properties to the optimal maximum likelihood networks. This makes sense because if $C_{\rm max}$ is large, the cutoff function has so little power that optimal method and $C_{\rm max}$ method are almost the same. Then, when we decrease $C_{\rm max}$, the properties of networks move towards those of BA networks. Notably, when $C_{\rm max}$ is small, the properties of $C_{\rm max}$ networks are quite similar to those of BA networks. All of above indicate that $C_{\rm max}$ method can actually link the BA method and optimal method. Panel (c) illustrates relative probability (relative to optimal maximum likelihood network) of $C_{\rm max}$ networks with $C_{\rm max}$ ranging from $15$ to $N$. Surprisingly, when $C_{\rm max}$ is neither very big nor very small, the relative probability will be higher than 1, which means the probability of truncated $C_{\rm max}$ network will be bigger than optimal networks. This suggests that the truncated approach provide a useful middle ground to generate networks ``between'' BA and the optimal growth networks.}
\label{figs/fig::2}
\end{figure*}

Figure $\ref{figs/fig5}$ provides representative realisations of the BA method, the optimal method, and the $C_{\rm max}$ method (for $C_{\rm max}$ between 20 and $N$). In Figure \ref{figs/fig5}, we can observe the structure of networks with different $C_{\rm max}$. It is easy to notice that when we decrease $C_{\rm max}$, the network evolve from a super-star structure to a scattered structure. In Figure $\ref{figs/fig::2}$, properties of $C_{\rm max}$ networks move from those of optimal networks to BA networks when we decrease $C_{\rm max}$ from $N$ to $20$, further supporting the qualitative observations of Figure $\ref{figs/fig5}$. Another interesting observation is that when $C_{\rm max}$ is rather small, networks can be very similar to BA networks, both in the structural sense shown in Figure $\ref{figs/fig5}$ and in the properties shown in Figure $\ref{figs/fig::2}$. The $C_{\rm max}$ method provides a link between the BA and optimal method via the parameter $C_{\rm max}$.

To identify whether a given network is more similar to BA or optimal network, the easiest and most direct way is to count the number (and size) of hubs. Hubs are really important in the network, because they are connected to many different nodes and therefore they share a high betweenness centrality. When we study the hubs, we can get a rough idea of the network. If there is only one hub (as with the super-star networks), it means that this network is highly concentrated, if there are many hubs, the network is more scattered.

Here we provide a working definition of these hubs. Naturally, the only judge of whether a node is a hub is the degree. We define the minimum degree of hubs as 
\begin{equation}
\nonumber
  \log(n_1)/\gamma + \theta
\end{equation}
 and any node with an equal or higher degree should be the hub. Here $\gamma$ is the degree exponent of the true asymptotic distribute, not that estimated from the data, and $\theta$ can be any value more than 1, with different $\theta$ the exact number of hubs may vary but it won't change the overall tendency. This is not the only definition of hubs, but alternative definitions will also yield the same tendency.

When we change the $C_{\rm max}$ value, we can see that the number of hubs also changes.  Figure $\ref{figs/fig::3}$ shows the tendency of the number of hubs. Decreasing $C_{\rm max}$ or increasing the size of network both lead to the increase of the number of hubs. This can be explained as follows. When the network grows, it is natural to generate more centres because as the population $N$ grows, nodes tend to (perhaps) gather into different groups and gather with different nodes instead of being all together. If $C_{\rm max}$ decreases,  the cutoff function has more influence and so it forces the richer nodes in the network to have fewer connections. However, as those richer nodes have fewer connections, their former neighbours need to connect elsewhere, those new connections increase the degree of other (relatively) low degree nodes,  increasing the prevalence of  hubs.

\begin{figure*}[!t]
\centering
\subfigure[Minimum Degree of hubs] {\includegraphics[width=8 cm]{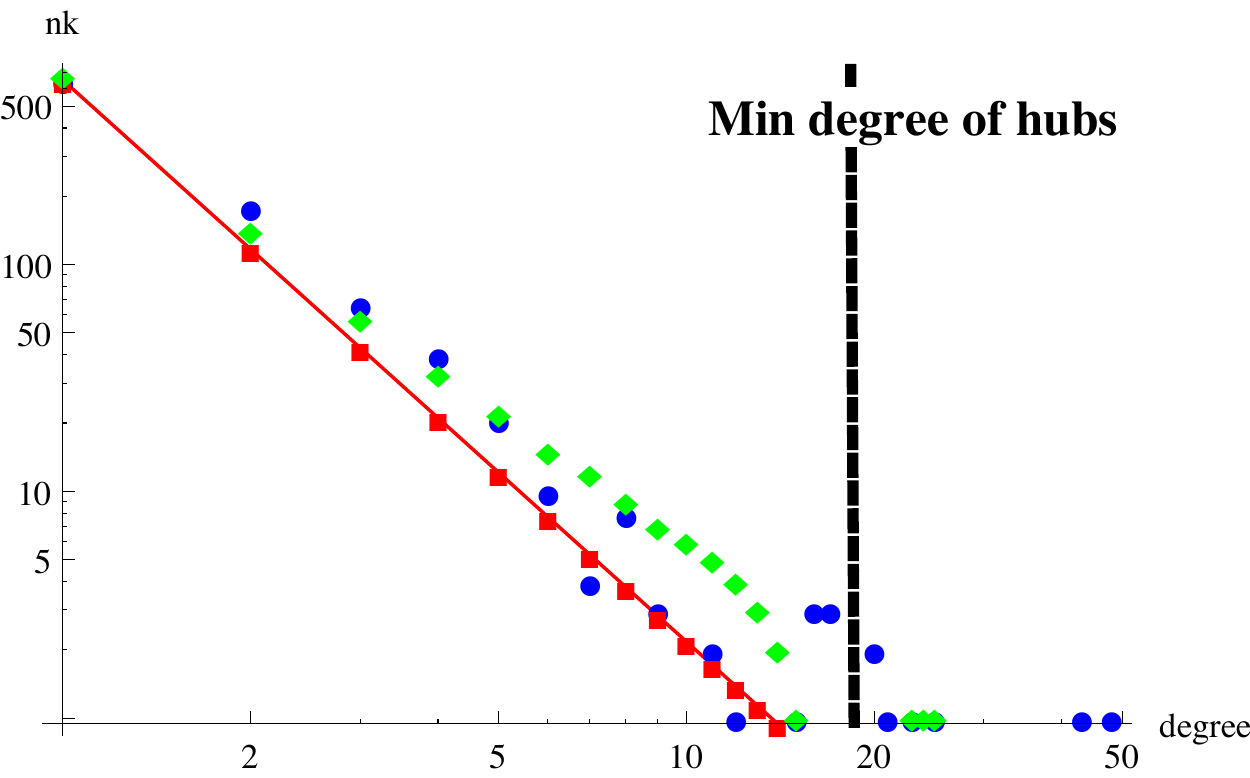}}
\subfigure[The number of hubs] {\includegraphics[width=8 cm]{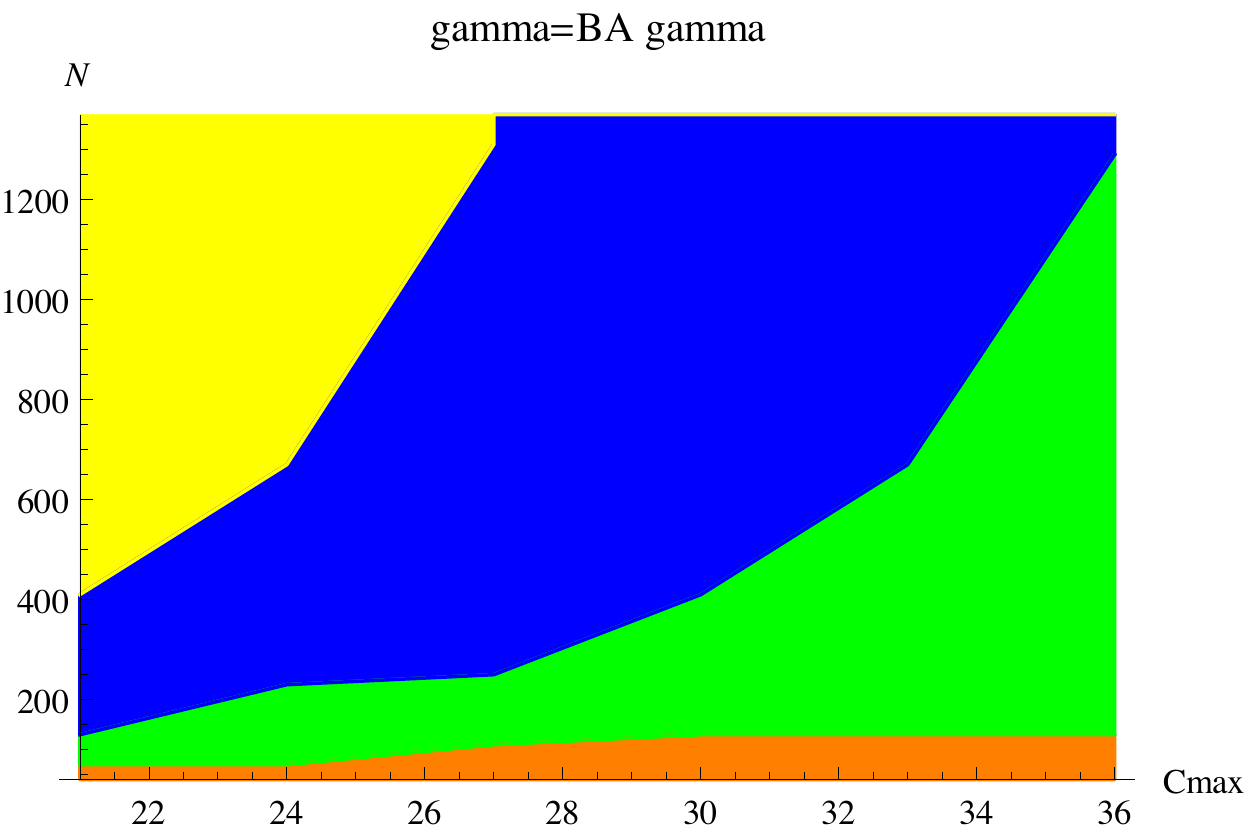}}
\caption{Panel (a) illustrates our constructive definition of hubs. With a pre-fixed $\gamma=2.47$ (corresponding to the estimated $\gamma$ in comparable BA networks), we draw a straight line in the degree distribution in log-log scale. The formula of the red line is: $\log(n_k)=\log(n_1)-\gamma \log(k)$, $k$ is degree and $n_k$ is the number of nodes with degree $k$. The intersection of this line and x axis is $\log(n_1)/\gamma$, since there is some variability at the end of the degree distribution, we add $\theta$ to exclude the tail. The green points show the degree distribution of $C_{\rm max}$ network with $C_{\rm max}=25$, and blue points shows BA network. Here we use $\theta=3$ and draw the minimum degree of hubs as the black line. From (a) we can see our definition successfully distinguish the hubs in the networks.
In (b), for $C_{\rm max}\in [21, 36]$ and $\gamma=2.47$ (the estimated $\gamma$ of BA method when minimum degree is 1), we compute realisations of $C_{\rm max}$ method and the number of hubs in the representative networks. The orange, green, blue, yellow areas indicate the number of hubs will be 1, 2, 3, 4 with particular $C_{\rm max}$ and $N$ in the representative areas. Note that if we fix $C_{\rm max}$, when we increase the size of networks, the number of hubs will increase, if we fix the $N$ and increase $C_{\rm max}$, the number of hubs will decrease.}
\label{figs/fig::3}
\end{figure*}


\section{Conclusion}
\label{conclusion}

Although we have derived four separate algorithms, we only examine three of them: the asymptotic scheme gives poor results if the seed network $G_s$ is small. Nonetheless, this  algorithm does provide insight into the asymptotic behaviour of the other methods.  Each of the three algorithms we present here provides a technique  to obtain random realisations of  networks consistent with a particular growth process. It has been argued that growing a network inherently biases the random sampling of the wider space of all networks consistent with a given degree distribution \cite{dC01}. While, in \cite{kJ13} we address the issue of random realisations from the space of all networks defined by a particular degree distribution, in this paper we propose a more narrowly defined growth algorithm. We demonstrate that  BA  is not the best way to {grow} networks consistent with a particular power-law degree distribution. In a sense, this extends the arguments of \cite{dC01} --- not only is random growth biasing one's selection from the space of all networks, BA is a biased selection \cite{skinny,mC05} from the space of randomly grown scale-free networks. Our algorithm provide a new approach to growing scale-free networks with an arbitrary degree exponent --- moreover, these networks exhibit a range of structural features beyond what one would expect from the BA. 

Of particular interest are the super-star networks that emerge from the optimal algorithm. These networks posses a unique structure not previously explored via standard growth processes. Recent work on explosive synchronisation in star networks \cite{zou1} demonstrates the importance of understanding this particular class of networks{.} Here, we see that super-star networks emerge via a natural and optimal growth process. This also provides a natural mechanism for the very large diameter and small exponent (i.e. $\gamma<2$) scale-free network observed in the real-world transmission of avian influenza \cite{birdflu} --- a large number of super-star hubs distributed geographically. Similar small exponent super-star networks have also been observed for networks of musical preference \cite{rL05} and sexual promiscuity \cite{fL01}. The algorithm we present provides a simple mechanism for generation of networks such as these. { Conversely, the single dominant ``super-star''
 can easily be forbidden by truncating (\ref{powerlaw}) with a harsher upper-bound dependent on $N$. Doing so produces networks with a distributed cluster of hubs.}
 

 \appendix
 \section{Expected degree}
 \label{edeg}
 \label{egamma}
 
 We provide an analytic expression for the expected degree exponent $\gamma$ of a preferential scale free network. The standard result \cite{rA02} holds that, for BA growth, $\gamma\rightarrow 3$ independent of $m$. However, this result is an asymptotic one concerning the tail of the distribution. A more useful statistic for what we are doing here is to estimate $\gamma$ from the entire distribution (\ref{powerlaw}). Doing so yields quite a different answer and we achieve excellent agreement between theory and computation. Moreover, this is a far more useful statistical measure for finite networks than the asymptotic result. After introducing our results we will briefly discuss the reasons behind the deviations from the results in \cite{rA02} in a little more detail.
  
 We perform a preferential attachment growth process to generate a scale-free network. At each stage we add a new node with $m$ new links. Let $k$ denote the degree of a node, and $N$ the number of nodes in the network. The degree distribution is assumed to converge to a power-law (for $k\geq m$) of the form $k^{-\gamma}$ and we obtain an exact implicit relationship for $\gamma$, $m$ and $N$. We verify this with  numerical calculations over several orders of magnitude. Although this expression is exact, it provides only an implicit expression for $\gamma(m)$. Nonetheless, we provide a reasonable guess as to the form of this curve and perform curve fitting to estimate the parameters of that curve --- demonstrating excellent agreement between numerical fit, theory, and simulation. 
 
Preferential attachment \cite{aB99} is the archetypal growth mechanism for scale-free networks. Asymptotically, under certain circumstances, such networks produce a degree distribution which converges asymptotically to a power law with exponent $3$. But this is not true in general, and it is not true for arbitrary finite networks generated along the way. In this note we derive straightforward analytic results for the expected exponent $\gamma$ of a scale free network with power law degree distribution $p(k)\propto k^{-\gamma}$. 

We assume that the network is grown with a Barab\'asi-Albert attachment process as described in \cite{aB99}. With each new node we add $m$ links and the growth process is terminated when the network has $N$ nodes. We make the approximation that the degree distribution of this finite networks follows a shifted power-law\footnote{That is, exactly a power law for degree $k\geq m$ where $m\geq 1$ is the number of edges added with each new node.} with some exponent $\gamma$.

Hence, a BA network with minimum degree $m$ will add exactly $m$ new links for each new node. The expected degree 
\begin{eqnarray}
\label{spower2}  E(k)&=&2m
\end{eqnarray} 
(since each link has two ends and contributed to the degree of two nodes). Conversely,  the probability that a node has degree $k$ is given by
\begin{eqnarray*}
P(k|\gamma,d) &=&\left\{\begin{array}{cc}
0 & k< m\\
\frac{k^{-\gamma}}{K(\gamma)} & k\geq m
\end{array}
\right.
\end{eqnarray*}
where the normalization factor $K(\gamma)$ is inconvenient. However
\begin{eqnarray*}
\zeta({\gamma} )&=& \left(\sum_{k=1}^{m-1} +\sum_{k=m}^\infty\right)k^{-\gamma} \\
&=&\sum_{k=1}^{m-1}k^{-\gamma}+K(\gamma)
\end{eqnarray*}
and hence it is easily computable.

\begin{figure}
\begin{center}
\begin{tabular}{cc}
\includegraphics[width=0.45\textwidth]{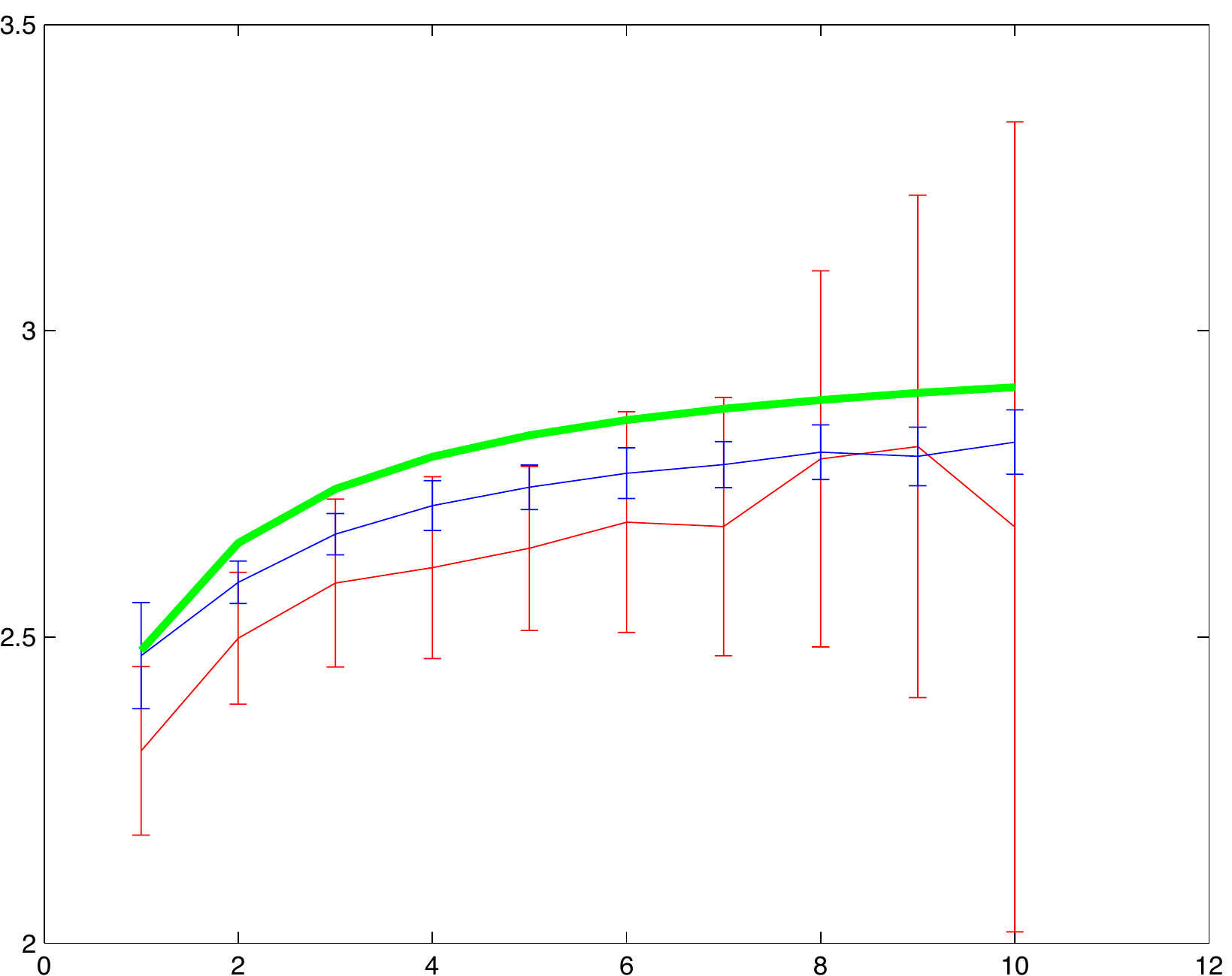} &
\includegraphics[width=0.45\textwidth]{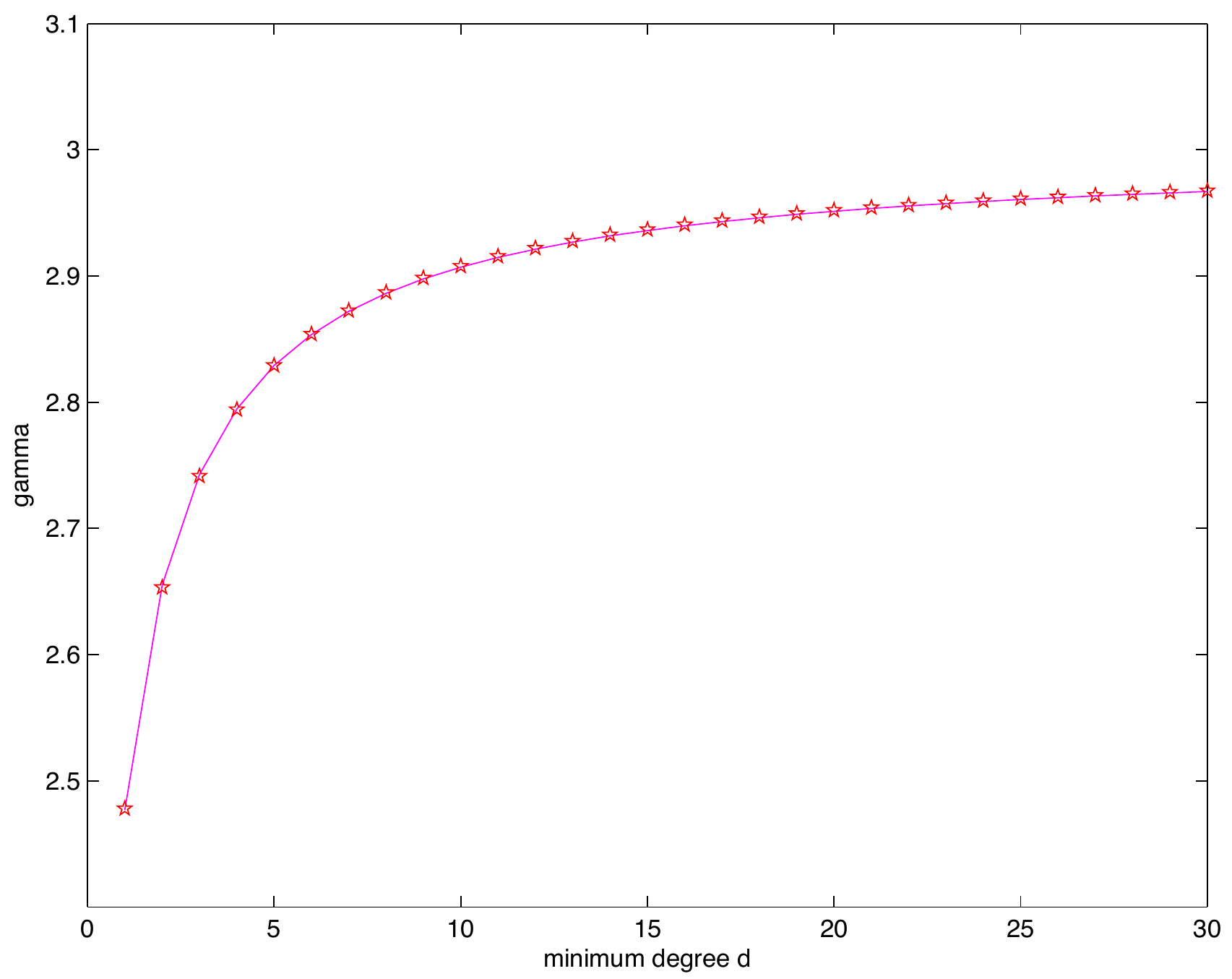}
\end{tabular}
\end{center}
\caption{Left panel: Expected values of $\gamma$ as a function of $m$ (Eqn.  (\ref{egam})) (heavy line) and estimated values of $\gamma$ from $30$ independent realisations of BA networks of size $N$ (mean $\pm$ standard deviation). We take $m\in[1,10]$ and $N=10^3$ (red), $10^4$ (blue), $10^5$ (green). Right panel: $\gamma$ as a function of $m$ computed via the solution of (\ref{mgamm2}) (stars) and estimated from a function fit of the form $\hat\gamma(m)= 3-(m+\alpha)^{-\beta}$. The best fit (obtained from a fit on $m\in[1,10]$) is then extrapolated over the domain. Parameter values are $\alpha= 0.9205$ and $\beta=0.9932$. }
\label{figs/fig_gd}
\end{figure}

The expected degree is
\begin{eqnarray}
\nonumber E(k)  &=&\sum_{k=1}^\infty kP(k|\gamma)\\
\label{egam}
 &=& \frac{\sum_{k=m}^\infty k^{1-\gamma}}{\zeta(\gamma)-\sum_{k=1}^{m-1}k^{-\gamma}}
 \end{eqnarray}
 Equating (\ref{spower2}) and (\ref{egam}), we have that the asymptotic value of $\gamma$ satisfies
 \begin{eqnarray}
 \label{mgamm}
\zeta(\gamma) &=& \sum_{k=1}^{m-1}k^{-\gamma}+\frac{1}{2m}\sum_{k=m}^\infty k^{1-\gamma}.
 \end{eqnarray}
Replacing the LHS of  (\ref{mgamm})  with the corresponding infinite sum and cancelling identical terms we obtain
\begin{eqnarray}
\label{mgamm2}
\sum_{k=m}^\infty (2m-k)k^{-\gamma}&=&0
\end{eqnarray}
 Solving   (\ref{mgamm2})  allows us to determine the expected value of $\gamma $ for the BA algorithm with a particular choice of minimum degree $m$. In particular, for $m=1$ we recover $2\zeta({\gamma})=E(k).$ 
 
 In Fig. \ref{figs/fig_gd} we illustrate the agreement between sample preferential attachment networks of various sizes and the prediction of (\ref{mgamm2}). The curve appears to be asymptotic to $\gamma=3$ and so we fit a function of the form $\hat\gamma(m)= 3-(m+\alpha)^{-\beta}$ to the solution of the series (\ref{mgamm2}). We obtain that
 \[\gamma(m)\approx 3-\frac{1}{(m+0.925)^{0.9932}}.\]
These results are required to explain expected degree distributions observed in Sec. \ref{stars} (Fig. \ref{newfig}), and in that case also show excellent agreement.

As noted above, this is not the same answer as that provided in the review of Albert and Barab\'asi \cite{rA02}.  The discrepancy arises from the methods used to estimate the exponent $\gamma$. Essentially, the standard maximum likelihood approach described by Newman \cite{mN05} imposes a minimum degree $x_{\rm min}$ and estimates $\gamma$ using a maximum likelihood expression \[\gamma = 1+n\left\slash\sum_{x_i>x_{\rm min}}\frac{x_i}{x_{\rm min}}\right..\]
This expression only becomes independent of $x_{\rm min}$ once $x_{\rm min}>\min_{i=1,\ldots, N}x_i$ and invokes Bayes' rule with a uniform prior on $\gamma$. Consequently, for $x_{\rm min}$ sufficiently large one observes that for preferential attachment $\gamma\rightarrow 3$ independent of $m$. We choose to maximise likelihood directly over the finite degree histogram.

The review paper \cite{rA02}  derives this asymptotic degree distribution in three different ways:
\begin{enumerate}
\item{\bf Continuum Theory:}
 Critically, the continuum approach assumes that $k_i$ is a continuous real variable when it is in fact a discrete random process. Moreover, the results only hold asymptotically. 
The continuity assumption holds only in the tail of the distribution --- avoiding the systematic bias away from the power law for low degree nodes.
\item{\bf Master-equation:} This approach actually obtains a slightly different expression for $P(k)$ which is dependent on $m$ and scales as the inverse of a cubic polynomial. Again, an approximation that is valid in the tail but not the head of the distribution. 
\item{\bf Rate-equation:} Similar to the previous approach, one will obtain an equivalent expression under the same assumption.
\end{enumerate}
Of these three approaches it is only the continuum approach which faithfully yields the claimed result that $P(k)\propto k^{-3}$ and this is only true under the assumption (true only in the tail) that $k_i$ is a continuous real variable. Albert and Barab\'asi acknowledge this in Sec. VII.C. of their review: ``these methods (the master- and rate- equation approaches), not using a continuum assumption, appear more suitable for obtaining exact results in more challenging network models.''

To estimate $\gamma$ in the case of pure preferential attachment (the BA process) there is no reason to insist on any choice other than $x_{\rm min}=m$ --- the entire distribution should be scale-free and to do otherwise unnecessarily favours the presumed asymptotic behaviour of the tail of a finite graph.  This is what we do here. We stress that we employ our estimate of $\gamma$ as nothing more than a descriptive statistic. For this purpose, it makes no sense to seek the asymptotic value which is independent of the structure we are trying to quantify. However, as the graphs we generate conform to the power-law over their entire range, the characterisation we produce here is also the correct exponent to describe that distribution.


\section{Generalised maximum likelihood growth}
\label{gmlg}

In the main text we focus on growth by adding a single new node or a single edge at each time step. However, the BA model of preferential attachment adds each new node and $m$ edges simultaneously. The resulting degree distribution is not (\ref{powerlaw}) and has the additional constraint that $p_k=0$ for $k<m$ where $m>1$. In particular $p_1=0$ and there is no chance of encountering a node with degree 1 (or any degree less than $m$). One possible computational expedient to overcome this problem is to replace $p_k=0$ with $p_k=\epsilon>0$ for $k<m$.  In this appendix we provide the exact likelihood expression --- extension of (\ref{q-link}) and (\ref{q-node}) for the case where one adds $m>1$ links simultaneously.

Let ${\cal D}_{m,N}$ denotes the sequence of  degrees of $m$ nodes chosen from among the nodes of the existing graph on $N$ nodes. The degree of $m$ nodes in $N$ are denoted as below, and $q_k$ here denotes the number of nodes in $N$ having degree $k$.
\begin{eqnarray}
{\cal D}_{m,N} & = &\{ \underbrace{k_1, k_1,\dots,k_1}_{q_{k_1}}, \underbrace{k_2, \dots, k_2}_{q_{k_2}},\dots, \underbrace{k_s, \dots, k_s}_{q_{k_s}} \}
\end{eqnarray}
where $k_1<k_2<\dots<k_s$ and $\sum_{i=1}^{s} q_{k_i} = m$. From the main text, we know that $Q_{node}$ can be described as
\begin{eqnarray} \nonumber
Q_{node}&= &\frac{P(G_{N+1})}{(N+1)\tilde{p}P(G_N)}.
\end{eqnarray}
However, the expression for connecting simultaneously to $m$ nodes is far less straightforward than the case  (\ref{q-node}) for $m=1$. In order to provide a more tractable formula, we need to change our notation. Let $l_1=k_1$. If $k_2=k_1+1$, then $l_1+1=k_2$, if not, let $l_2=k_2$. Hence we can rewrite ${\cal D}_{m,N}$ as
\begin{eqnarray}
\label{matr}
 \text{\cal D}_{m,N} & = &
  \{ l_1,  l_1+1,  \cdots,  l_1+b_1,  l_2,  l_2+1,  \cdots,  l_2+b_2, 
    \cdots 
    l_t,  l_t+1,  \cdots,  l_t+b_t, \}
\end{eqnarray}
where $k_1=l_1$, $\dots$, $k_{b_1+1}=l_1+b_1$ and $k_{(\sum_{i=1}^{t-1}b_i )+t-1}=l_t,\dots,k_m=l_t+b_t$.
The sequence on the right hand side of (\ref{matr}) is a complete list of available node degrees (not counting multiplicities) parameterised under the $l_i$'s and $b_j$'s.

Use this new notation, the same counting argument as described in the main text for $m=1$ now yields:
\begin{eqnarray}
\label{equ::appB}
Q_{{\rm node}-m}(\cal{D}_{m,N}) &=&
\left\{\begin{array}{cc}
\begin{array}{c}
\frac{p_m}{n_m+1}\frac{1}{\tilde{p}}\prod_{i=1}^{t}\frac{p_{l_i+b_i+1}^{q_{l_i+b_i}}}{p_{l_i}^{q_{l_i}}}\frac{n_{l_i+b_i+1}!}{(n_{l_i+b_i+1}+q_{l_i+b_i})!}\frac{n_{l_i}!}{(n_{l_i}-q_{l_i})!}\times\ldots \\
\;\;\;\;\;\;\;\;\prod_{j=1}^{b_i} p_{l_i+j}^{{q_{l_i+j-1}-q_{l_i+j}}} \frac{n_{l_i+j}!}{(n_{l_i+j}-q_{l_i+j}+q_{l_i+j-1})!}
\end{array}
  &
l_1> m\\
&\\
\begin{array}{c}
\frac{p_m}{(n_m+1-q_{m})}\frac{1}{\tilde{p}}\prod_{i=1}^{t}\frac{p_{l_i+b_i+1}^{q_{l_i+b_i}}}{p_{l_i}^{q_{l_i}}}\frac{n_{l_i+b_i+1}!}{(n_{l_i+b_i+1}+q_{l_i+b_i})!}\frac{n_{l_i}!}{(n_{l_i}-q_{l_i})!}\times\ldots\\
\;\;\;\;\;\;\;\;\prod_{j=1}^{b_i} p_{l_i+j}^{{q_{l_i+j-1}-q_{l_i+j}}} \frac{n_{l_i+j}!}{(n_{l_i+j}-q_{l_i+j}+q_{l_i+j-1})!}
\end{array}
 &
l_1=m
\end{array}\right.
\end{eqnarray}
There are several things we need to notice. First, when $m=1$, formula ($\ref{equ::appB}$) is the same as the formula (\ref{q-node}) given in the main text. Second, when we add a new edge, $Q_{edge-j}$ is identical to (\ref{q-link}). Finally, when a new node added, it connects to the $m$ nodes at the same time. The following is not allowed: if $n_k=0$, and $n_{k-1}>0$, the new node is connected to a node with degree $k-1$ and then connected to the same node which now has degree $k$. So if $n_k = 0$, then $k\notin {\cal D}_{m,N}$.

 \section*{Acknowledgements}
 
MS  is funded by the Australian Research Council via a Future Fellowship (FT110100896) and Discovery Project (DP140100203). YL is supported by the UWA-USTC research training programme. Mathematica and MATLAB implementations of the algorithms described in this paper are available from the first author.


%

\end{document}